\documentclass[12pt]{article}
\usepackage{epsfig}
\newcommand{\mysection}{\setcounter{equation}{0}\section}

\def\beq{\begin{equation}}
\def\eeq{\end{equation}}
\def\beqa{\begin{eqnarray}}
\def\eeqa{\end{eqnarray}}

\newlength{\dinwidth} \newlength{\dinmargin}
\begin{document}
\begin {flushright}
Cavendish-HEP-04/01\\
LBNL-54251\\
\end {flushright} 
\vspace{3mm}
\begin{center}
{\Large \bf Threshold corrections 
in bottom and charm quark hadroproduction at next-to-next-to-leading order}
\end{center}
\vspace{2mm}
\begin{center}
{\large Nikolaos Kidonakis$^a$ and Ramona Vogt$^b$}\\
\vspace{2mm}
$^a${\it Cavendish Laboratory, University of Cambridge,\\
Madingley Road, Cambridge CB3 0HE, UK}\\
\vspace{2.5mm}
$^b${\it Nuclear Science Division,\\
    Lawrence Berkeley National Laboratory, Berkeley, CA 94720, USA \\
  and \\
  Physics Department,\\
    University of California at Davis, Davis, CA 95616, USA}
\end{center}

\begin{abstract}
We calculate threshold soft-gluon corrections
to total cross sections  and transverse momentum distributions
for bottom and charm quark production in fixed-target experiments, considering
both $pp$ and $\pi^- p$ interactions.
We investigate the quality of the near threshold soft-gluon approximation
at next-to-leading order (NLO) and calculate next-to-next-to-leading 
order (NNLO) corrections through next-to-next-to-next-to-leading-logarithmic 
(NNNLL) accuracy, including some virtual terms.
We find that the NNLO threshold corrections reduce the factorization 
and renormalization scale dependence of the cross sections.

\end{abstract}

\thispagestyle{empty}  \newpage  \setcounter{page}{2}

\mysection{Introduction}

The latest calculations for heavy quark hadroproduction 
have included next-to-next-to-leading-order (NNLO) soft-gluon 
corrections to the double differential cross section \cite{NKtop,KLMV,
KLMVb,KLMV_cc,KVtop} 
from threshold resummation techniques \cite{KS,KOSr,LOS}. 
These resummations are a consequence of the
factorization properties of QCD which separate cross sections into
universal nonperturbative parton densities and a perturbatively-calculable
partonic cross section. 
Near threshold there is limited phase space for the emission of real
gluons so that soft-gluon corrections dominate the cross section.
These Sudakov corrections have the form of logarithmic
plus distributions, singular at partonic threshold.
Threshold resummation techniques organize these singular distributions to all
orders, thus extending the reach of QCD into the near-threshold region. 

Calculations of bottom and especially charm production are still 
not under solid theoretical control.  
A good understanding of the bottom cross section is
important for HERA-B, where $b$-quark production is near threshold.
The charm cross section is of 
particular interest for heavy ion physics.  
Although many future heavy ion experiments will be at high collider energies,
$\sqrt{S} \geq 130$ GeV, 
some of the current and future experiments, like those
at the SPS, are in the near-threshold region.  The NA60
experiment will take heavy ion data at $\sqrt{S} = 17.3$
GeV and $pA$ data at $\sqrt{S} = 29.1$ GeV.  
A new facility is being built at
the GSI \cite{gsi} that will measure charm near threshold with $\sqrt{S}=6.98$
GeV.

Because the charm quark mass is a few times $\Lambda_{\rm QCD}$, 
it is generally treated as a
heavy quark in perturbative QCD calculations.  However, its relative
lightness results in a rather strong mass and scale dependence for the total 
cross section.  
There is also a rather broad spread in the charm 
production cross section data at fixed target energies.  

In this paper, we increase the accuracy of previous soft-gluon
calculations for bottom and charm production.
The soft corrections that we calculate take the form of logarithms,
$[\ln^l(x_{\rm th})/x_{\rm th}]_+$, with $l \le 2n-1$ for the 
order $\alpha_s^n$ corrections, where $x_{\rm th}$ is a kinematical
variable that measures distance from threshold and goes to zero
at threshold.
NNLO calculations ($n=2$) for bottom and charm quark production 
have so far been done
through next-to-next-to-leading-logarithmic (NNLL) accuracy, i.e. for the
scale-independent terms, including leading logarithms (LL) with $l=3$, 
next-to-leading logarithms (NLL) with $l=2$, and NNLL with 
$l=1$ \cite{NKtop,KLMV,KLMVb,KLMV_cc}. 
In Refs.~\cite{KLMV,KLMVb,KLMV_cc}, heavy quark  cross sections were studied 
in both single-particle-inclusive (1PI) and pair-invariant-mass (PIM) 
kinematics. Important differences  between the two kinematics choices
were found in both the parton-level and hadron-level cross sections, even
near threshold. 
Thus subleading, i.e. beyond NNLL, contributions can still have
an impact on the cross section.  Their inclusion is clearly
needed to bring the calculation under further theoretical control.
The subleading terms indeed minimize the kinematics dependence of the top quark
production cross section \cite{KVtop}.

In Refs.~\cite{NKtop,KLMV} we studied top and bottom quark production
at NNLO-NNLL. In Ref.~\cite{KLMV_cc} we studied charm quark production
at NNLO-NNLL. More recently we extended our calculations for
top quark production \cite{KVtop}, using
the methods and results of Ref.~\cite{NKNNLO}, 
to include additional subleading NNLO soft corrections,
including next-to-next-to-next-to-leading logarithms (NNNLL), with $n=2$, 
$l=0$, as well as some virtual $\delta(x_{\rm th})$ corrections.
We showed in Ref. \cite{KVtop} that the subleading corrections do 
indeed bring the 1PI and PIM results into agreement near threshold for both the
$q {\overline q} \rightarrow Q {\overline Q}$ and the
$gg \rightarrow Q {\overline Q}$ channels
while the discrepancies away 
from threshold are also diminished, especially in the $gg$ channel.
Thus the threshold region is brought under better theoretical control.
In this paper, we apply these new terms to bottom and charm quark production.
 
In the following section, we briefly discuss the differences 
between the 1PI and PIM
kinematics choices and then investigate which choice is a better
approximation to the full NLO result. We find that 1PI kinematics
is a far better choice.  Thus we only calculate the cross section in 
this kinematics in the remainder of the paper. 

Section 3 discusses bottom quark production in $\pi^- p$
and $pp$ collisions, particularly at HERA-B.
We present the hadronic total cross sections for a range
of energies near threshold.  We also calculate the transverse
momentum distributions for bottom production at HERA-B.
In Section 4 we study charm production in $\pi^- p$
and $pp$ collisions.
Our studies focus on the kinematics of the proposed GSI facility and
the CERN SPS proton and ion fixed target programs.
We conclude with a summary in Section 5.

\mysection{Singular distributions and kinematics dependence}

As we discussed in the introduction, the soft-gluon
corrections that we calculate
take the form of logarithms,
$[\ln^l(x_{\rm th})/x_{\rm th}]_+$,  where $x_{\rm th}$ is a kinematical
variable that measures the distance from partonic threshold.
The exact definition of $x_{\rm th}$ depends on the kinematics
we use for the calculation of the cross section.

We study the partonic process $ij \rightarrow Q {\overline Q}$
where $Q$ is the produced heavy quark and $ij$ can be either
$q \overline q$ or $gg$.  
A more detailed discussion of the kinematics can be found
in Ref.~\cite{KLMV}. 

In 1PI kinematics, a single heavy quark is identified so that
\beq
i(p_a) + j(p_b) \longrightarrow Q(p_1) + X[{\overline Q}](p_2)
\eeq
where $Q$ is the identified bottom or charm quark of mass $m$
and $X[{\overline Q}]$ is the remaining final state that contains
the heavy antiquark ${\overline Q}$.
We define the kinematical invariants 
$s=(p_a+p_b)^2$, $t_1=(p_b-p_1)^2-m^2$, $u_1=(p_a-p_1)^2-m^2$
and $s_4=s+t_1+u_1$. At threshold, $s_4 \rightarrow 0$
and the soft corrections appear as $[\ln^l(s_4/m^2)/s_4]_+$.
We note that the virtual corrections multiply $\delta(s_4)$.

In PIM kinematics, we have instead
\beq
i(p_a) + j(p_b) \longrightarrow Q{\overline Q}(p) + X(k) \, .
\label{PIM}
\eeq
At partonic threshold, $s=M^2$, $M^2$ is the pair mass squared, 
$t_1=-(M^2/2)(1-\beta_M \cos\theta)$ and
$u_1=-(M^2/2)(1+\beta_M \cos\theta)$
where $\beta_M=\sqrt{1-4m^2/M^2}$ and $\theta$ is the scattering
angle in the parton-parton center-of-mass frame.
The soft corrections appear as $[\ln^l(1-z)/(1-z)]_+$
with $z=M^2/s \rightarrow 1$ at threshold.
We note that the virtual corrections multiply $\delta(1-z)$.

The effect of the soft-gluon corrections varies depending
on the kinematics choice because of uncalculated
subleading terms. Thus the quality of the soft-gluon
approximation, the extent to which the soft-gluon corrections
approximate the full corrections
at a given order in $\alpha_s$, 
depends on the kinematics choice even though the full corrections 
do not depend on the kinematics.

\begin{figure}[htpb]
\setlength{\epsfxsize=1.0\textwidth}
\setlength{\epsfysize=0.5\textheight}
\centerline{\epsffile{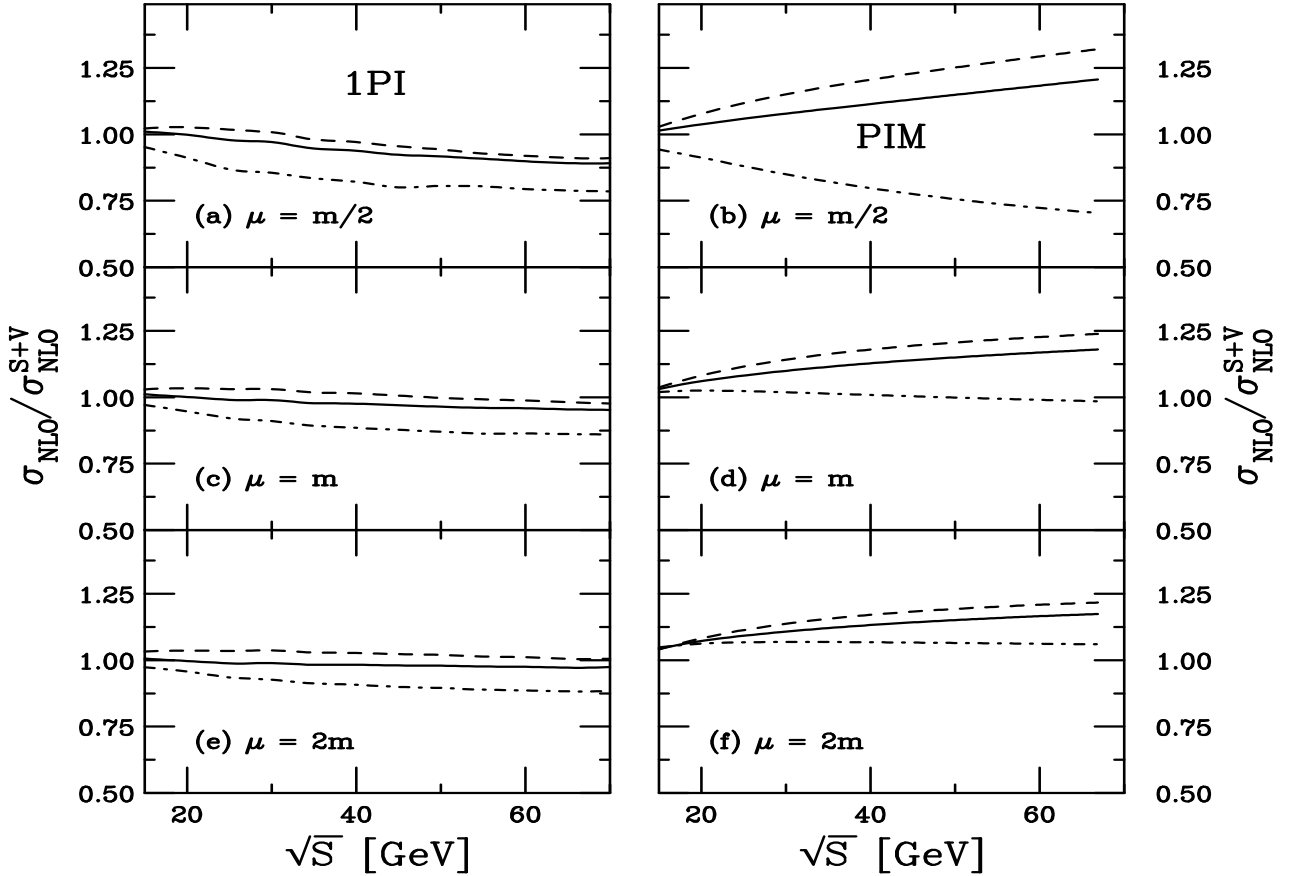}}
\caption[]{The ratios of the NLO exact, $\sigma_{\rm NLO}$ over the
NLO soft-plus-virtual, $\sigma_{\rm NLO}^{\rm S+V}$,
cross sections for bottom quark production with $m = 4.75$ GeV are
shown for the $gg$ (dashed) and $q{\overline q}$
(dot-dashed) channels separately, along with their sum (solid).
}
\label{botK1pipim} 
\end{figure}

We thus first test the soft-gluon approximation in 1PI and PIM
kinematics at NLO. In Fig.~\ref{botK1pipim} we present the ratios of
the exact NLO cross sections, $\sigma_{\rm NLO}$, to the approximate 
soft-plus virtual NLO cross section, $\sigma_{\rm NLO}^{\rm S+V}$,
(the Born result plus the ${\cal O}(\alpha_s^3)$ soft and virtual 
corrections)
for bottom quark production in $pp$ collisions over a range of
$\sqrt{S}$ near bottom production threshold.
In our calculations we use the MRST2002 NNLO \cite{mrst2002} parton densities.
We show the ratios in both 1PI and PIM kinematics for each of the two
partonic channels separately as well as for their sum for three scale
choices: $\mu=m/2,m$, and $2m$.
Note that the factorization and renormalization 
scales are set equal and are both denoted by $\mu$.

\begin{figure}[htpb]
\setlength{\epsfxsize=1.0\textwidth}
\setlength{\epsfysize=0.38\textheight}
\centerline{\epsffile{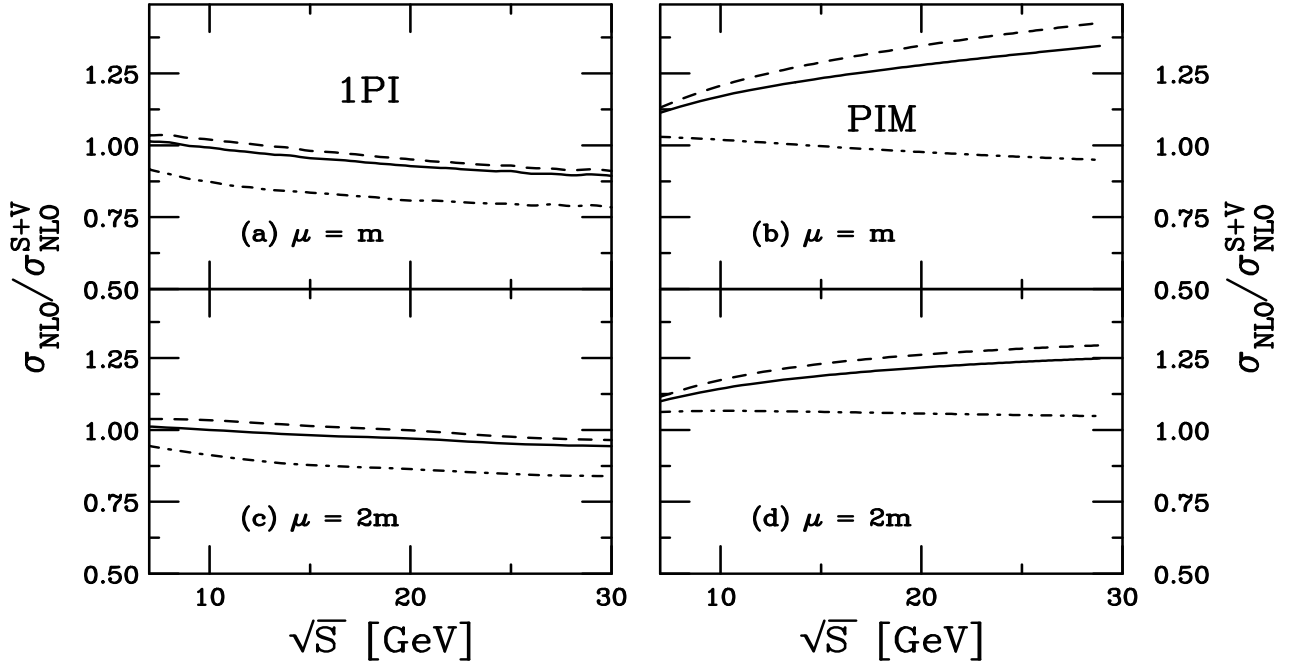}}
\caption[]{The ratios of the NLO exact, $\sigma_{\rm NLO}$ over the
NLO soft-plus-virtual, $\sigma_{\rm NLO}^{\rm S+V}$,
cross sections for charm quark production with $m = 1.5$ GeV are
shown for the $gg$ (dashed) and $q{\overline q}$
(dot-dashed) channels separately, along with their sum (solid).
}
\label{chmK1pipim} 
\end{figure}

It is clear from Fig.~\ref{botK1pipim} that the soft-plus-virtual
result is a much better approximation to the exact NLO total cross section
in 1PI kinematics than in PIM kinematics over the entire
range of energies and scales shown. The same conclusion is reached for charm
production, shown in Fig.~\ref{chmK1pipim} for
$pp$ collisions with $\mu=m$ and $2m$. Since the 1PI kinematics choice is a
better approximation to the exact NLO cross section, here we choose to work 
in this kinematics only.
 
One-particle inclusive kinematics is a better approximation to the total
$b \overline b$ and $c \overline c$ production cross sections
because the $gg$ channel is dominant.  For $\mu \geq m$, PIM kinematics gives a
better approximation to the NLO $q \overline q$ cross section, not unexpected
due to the simple color structure of this $s$-channel process.  However,
with its more complex color structure, the $gg$ channel is more amenable to the
1PI kinematics choice.  We also found that, when the $gg$ channel dominates
production, the 1PI scale dependence remains small while the
PIM kinematics choice has a large scale dependence, larger than the exact NLO
\cite{KVtop}. 

\begin{figure}[htpb]
\setlength{\epsfxsize=0.7\textwidth}
\setlength{\epsfysize=0.4\textheight}
\centerline{\epsffile{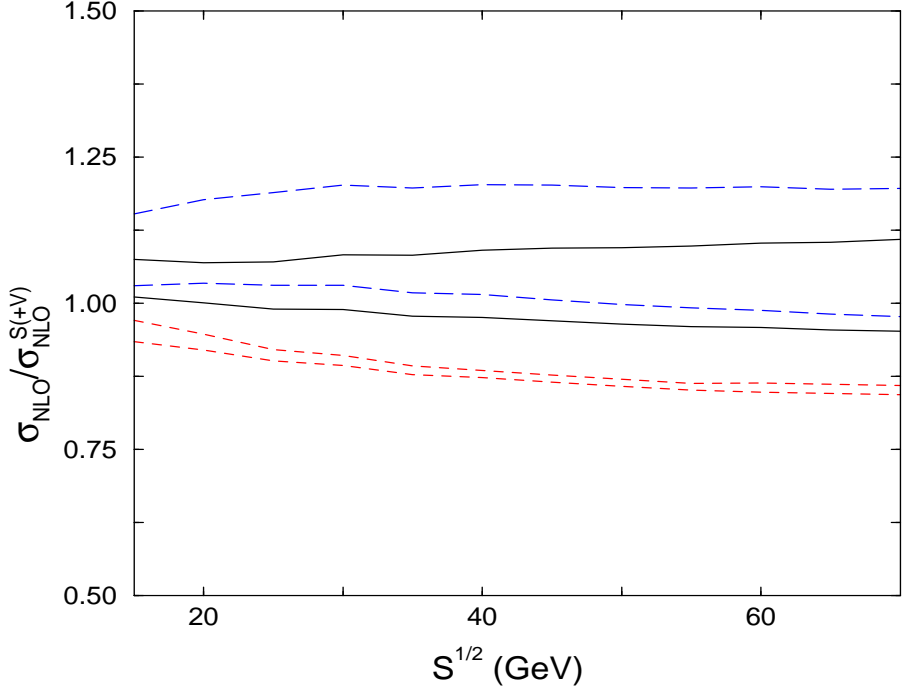}}
\caption[]{The ratio of $\sigma_{\rm NLO}/\sigma_{\rm NLO}^{\rm
S}$, with the soft gluon terms alone, and $\sigma_{\rm NLO}/\sigma_{\rm
NLO}^{\rm S+V}$, including the virtual terms, are shown for 
bottom quark production at $\mu=m = 4.75$ GeV.
We show the results for the $gg$ channel (long-dashed) 
with (lower) and without the virtual terms (upper);
the $q {\overline q}$ channel (dashed) with
(upper) and without the virtual terms (lower); and their sum
(solid) with (lower) and without the virtual terms (upper).
}
\label{KNLL} 
\end{figure}

Finally, we investigate the validity of the 
threshold approximation if only the soft terms are included and the 
virtual ones are not.
Figure \ref{KNLL} compares the ratios, $\sigma_{\rm NLO}/\sigma_{\rm NLO}^{\rm
S}$, with the soft gluon terms alone, to $\sigma_{\rm NLO}/\sigma_{\rm
NLO}^{\rm S+V}$ for bottom quark production in 1PI kinematics
with $\mu=m=4.75$ GeV. 
We see that even though the ratios with the NLO soft gluon approximation 
alone are somewhat farther from one, they are still a rather good
estimate of the exact NLO result.

In the following sections we present cross sections
for bottom and charm quark production. 
The LO and NLO results are exact. The NNLO results are approximate
and include soft-gluon contributions only.
At NNLO we  give results both for NNLL accuracy, i.e. for the
scale-independent terms, including
all $\ln^3(s_4/m^2)/s_4$, $\ln^2(s_4/m^2)/s_4$, 
and $\ln(s_4/m^2)/s_4$ terms,  and for  
NNNLL+$\zeta$ accuracy,
which includes, in addition to the NNLL terms, all $1/s_4$ terms (NNNLL)
as well as some $\zeta$ terms in the virtual corrections.
For full details, see Ref. \cite{KVtop}. We note that the 
NNLO-NNNLL+$\zeta$ result includes all NNLO soft and virtual terms 
proportional to the factorization and renormalization scales, while the 
NNLO-NNLL calculation includes in the virtual contribution only 
scale-dependent terms proportional to the squares of scale-dependent 
logarithms.

\mysection{Bottom quark production}

The main difficulty in comparing the calculated $Q \overline Q$ 
cross sections to
data lies in the fact that charm and bottom quarks hadronize before decaying.
Top quark production is much cleaner in this respect since the top quark decays
before it can hadronize and the measured cross sections can be compared
directly to theory.  No attempt has been made to correct $B$ meson measurements
to $b$ quark cross sections here.

\subsection{$b$-quark production in $\pi^- p$ interactions}

\begin{figure}[htpb]
\setlength{\epsfxsize=0.6\textwidth}
\setlength{\epsfysize=0.6\textheight}
\centerline{\epsffile{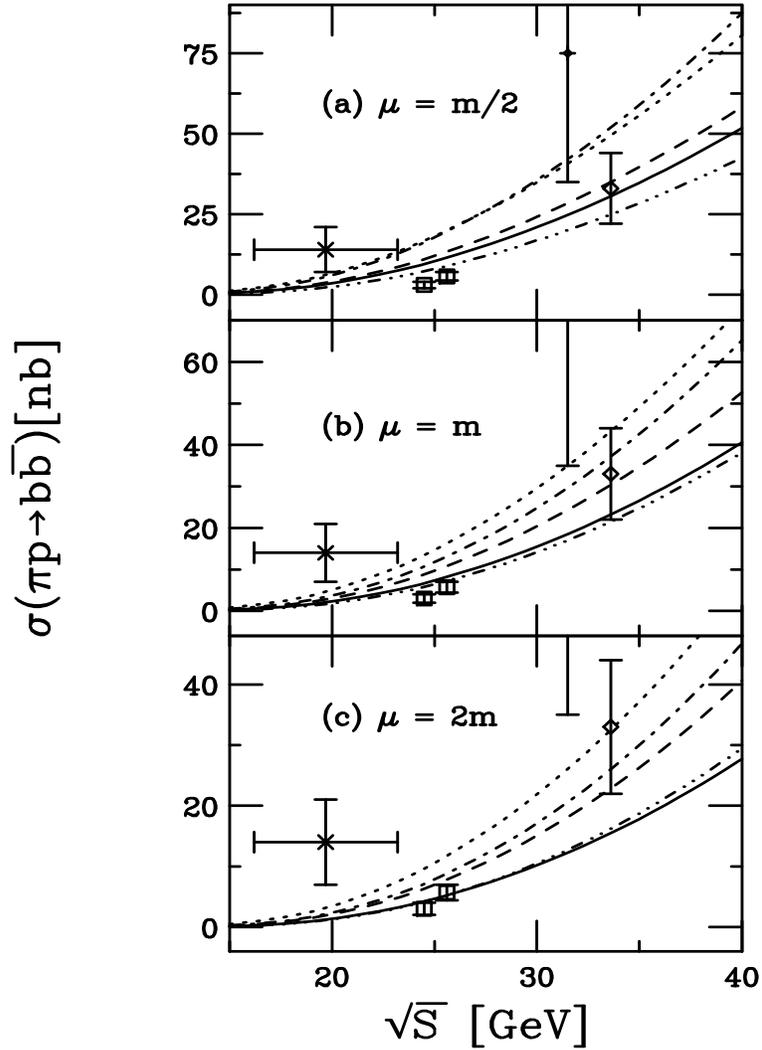}}
\caption[]{The energy dependence of $b{\bar b}$
production in $\pi^- p$ collisions with (a) $\mu = m/2$, (b) $\mu =m$ and (c)
$\mu = 2m$, calculated with the GRV98 HO proton densities and the GRS pion
densities.  We show the NLO (solid) and 1PI
NNLO-NNLL (dot-dashed) results at $m = 4.75$ GeV.  The 1PI NNLO-NNNLL$+\zeta$
results are shown for $m = 4.75$ (dashed), 4.5 (dotted) and 5
(dot-dot-dot-dashed) GeV.
}
\label{f1} 
\end{figure}

There is not much data on bottom quark hadroproduction at fixed-target
energies.  Some of the earliest data were from $\pi^- A$ interactions where $A$
is a nucleus.  A linear $A$ dependence was typically assumed.
The data \cite{na10,wa78,wa92,e653,e672} 
are shown in Fig.~\ref{f1} along with our 
calculations.  The NA10 point at $\sqrt{S} = 19.7$ GeV is a compilation of
data taken at 140, 194 and 286 GeV on tungsten targets measuring trimuons
\cite{na10}.  They quote a $B \overline B$ cross section essentially
independent of energy.  The $x_F$ distribution of the $B$ mesons was assumed to
be uniform around $x_F = 0$ with a cross section proportional to 
$(1 - |x_F|)^3$.  Assumptions regarding the shape of the distribution and the
decay branching ratios almost certainly overestimate the cross section.  The
other data is analyzed using the $B$ decay to $J/\psi$.
The WA78 data, taken with 320 GeV $\pi^-$ beams on a uranium target,
used several hypotheses regarding production models and $B^0-\overline B^0$
mixing angles.  The data point shown assumes the largest $\langle x_F \rangle$
and the maximal mixing.  It is the smallest cross section reported in
Ref.~\cite{wa78}.  The WA92 data were taken at 350 GeV on a copper target
\cite{wa92}.  The E653 data was taken with a 600 GeV beam on emulsion
\cite{e653} while the E672 data at 515 GeV was taken on a beryllium target
\cite{e672}.  Only this latter point used a NLO calculation to extrapolate from
$x_F > 0$, where the data was taken, to all $x_F$.  All these data rely on
relatively few events and some of the assumptions should be taken with a grain
of salt.

Figure~\ref{f1} presents the energy dependence of the production cross 
sections as functions of both bottom quark mass and scale.  We use the GRV98 HO
proton parton densities \cite{grv98} with the GRS pion densities \cite{grs}.  
The GRS pion densities, produced in 1999, are the most recent set of pion
densities and are compatible with the GRV98 densities.  We do not use other
proton densities because there are no recent equivalent pion sets.

Since the bottom quark mass is relatively large, we vary the scale between
$m/2$ and $2m$ in Fig.~\ref{f1}.  For our central value of the bottom quark
mass, $m = 4.75$ GeV, we present the exact NLO cross section (solid curve), 
the 1PI NNLO-NNLL cross section (dot-dashed) and the 1PI NNLO-NNNLL$+\zeta$ 
cross section (dashed).  We also show the 1PI NNLO-NNNLL$+\zeta$ cross sections
for $m = 4.5$ GeV (dotted) and 5 GeV (dot-dot-dot-dashed).  Because we focus 
on the NNLO-NNNLL$+\zeta$ results here, we do not present the NLO and NNLO-NNLL
cross sections for the lower and upper mass values in Fig.~\ref{f1}.

\begin{figure}[htpb] 
\setlength{\epsfxsize=0.6\textwidth}
\setlength{\epsfysize=0.6\textheight}
\centerline{\epsffile{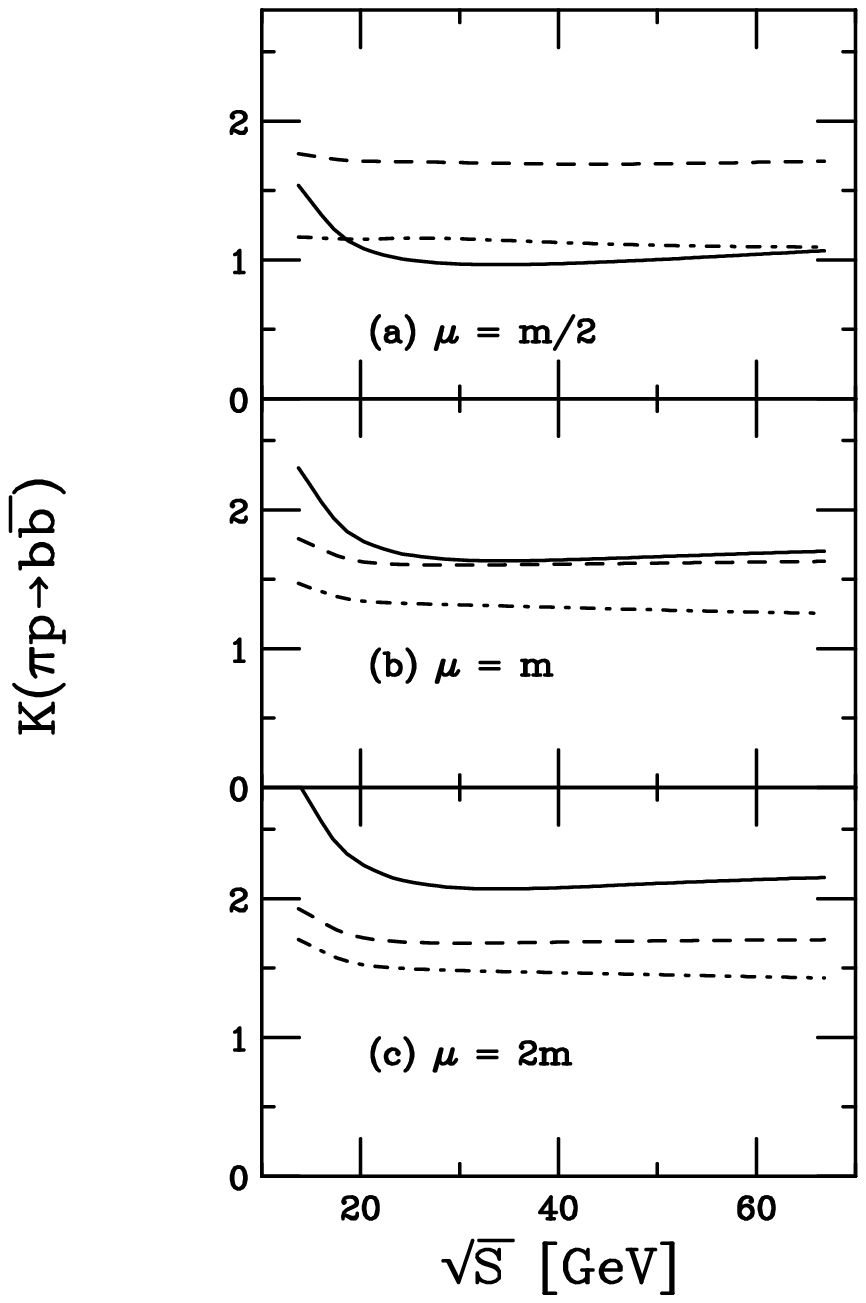}}
\caption[]{The $K$-factors for $b{\bar b}$
production in $\pi^- p$ collisions with $m=4.75$ GeV. 
We present $K_0^{(1)}$ (solid), $K^{(2)}$ (dashed) and $K^{(2)}_{\rm sub}$ 
(dot-dashed) results for (a) $\mu=m/2$, (b) $\mu = m$ and (c) $\mu = 2m$.
}
\label{f2} 
\end{figure}

It is difficult to quantitatively compare the calculations to the data because
the data are not really compatible with each other.  Therefore our discussion
is only on a qualitative level.  Note that, at $m = 4.75$ GeV, the 1PI
NNLO-NNNLL$+\zeta$ result is intermediate to the exact NLO and the 1PI
NNLO-NNLL cross sections.  The subleading terms reduce the overall NNLO
corrections, resulting in a value near the average of the 1PI and PIM NNLO-NNLL
cross sections.  The $m=4.5$ GeV 1PI NNLO-NNNLL$+\zeta$ cross section
is typically equivalent
to ($\mu = m/2$) or larger than ($\mu \geq m$) the $m=4.75$ GeV 1PI NNLO-NNLL 
cross section while the 5 GeV result is similar to the exact NLO
cross section for  $m=4.75$ GeV.  
Changing the bottom quark mass between
4.5 and 5 GeV changes the 1PI NNLO-NNNLL$+\zeta$ cross sections by a factor of
two. 

Near threshold, $\pi^- p \rightarrow b \overline b$ production is dominated
by the $q \overline q$ channel, due to the valence quark-valence antiquark
contribution at large momentum fractions \cite{SVchm}.  
The $q \overline q$ channel gives
the largest contribution to the $b \overline b$ cross section for $\sqrt{S}
\leq 40$ GeV, the range shown in Fig.~\ref{f1}.

We now discuss the convergence properties of the subleading terms.
Various definitions of the `first-order $K$ factor',
$K^{(1)} = \sigma_{\rm NLO}/\sigma_{\rm LO}$, were discussed in
Ref.~\cite{RVkfac}.  In Ref.~\cite{KLMVb,KLMV_cc}, we compared two of these
definitions, $K_0^{(1)}$, where $\sigma_{\rm LO}$ was 
calculated with NLO parton
densities and a two-loop evaluation of $\alpha_s$, and $K_2^{(1)}$, where 
$\sigma_{\rm LO}$ was calculated with LO parton densities and a one-loop
evaluation of $\alpha_s$.  In both cases, the Born and ${\cal O}(\alpha_s^3)$
contributions to $\sigma_{\rm NLO}$ were calculated with NLO parton densities
and two-loop evaluations of $\alpha_s$.  The first definition, $K_0^{(1)}$,
indicates the convergence of terms in a fixed-order calculation while the
second, $K_2^{(1)}$, indicates the convergence of the hadronic calculation
towards a result.  Since a NNLO set of parton densities has recently become
available, in this paper, we will use both NLO and NNLO parton
densities to calculate $K_0^{(1)}$.  
Thus when the NNLO densities are used, $\sigma_{\rm
LO}$ in $K_0^{(1)}$ is calculated with the NNLO parton densities and a
three-loop evaluation of $\alpha_s$.  The definition of $K_0^{(1)}$ then
indicates the convergence of terms in a fixed-order calculation for a given set
of parton densities.  We compare $K_0^{(1)}$ to the 1PI NNLO $K$ factors, 
$K^{(2)} = \sigma_{\rm NNLO-NNLL}/\sigma_{\rm NLO}$ and  $K_{\rm sub}^{(2)} =
\sigma_{{\rm NNLO-NNNLL}+\zeta}/\sigma_{\rm NLO}$; the latter
includes the new subleading terms.  In each case, the NNLO $K$ factors are
either calculated with NLO or NNLO parton densities and a two-loop or
three-loop evaluation of $\alpha_s$ at each order.  If $K^{(2)}$ or 
$K_{\rm sub}^{(2)}$ is less than $K_0^{(1)}$, then convergence of the
perturbative expansion is indicated.

The results for $\pi^- p \rightarrow b \overline b$ production are shown in
Fig.~\ref{f2} as a function of energy.  
We present the $K$ factors for $m = 4.75$ GeV only since the
results are very similar for the other masses.  All the $K$ factors are
calculated with the GRV98 HO proton parton densities and the GRS pion parton
densities.  Note that $K_0^{(1)}$ has the strongest energy dependence with a
minimum around $\sqrt{S} = 20$ GeV, higher at lower $\sqrt{S}$ and slowly
increasing with $\sqrt{S}$ above 20 GeV.  We find $K_0^{(1)} \approx 1$ 
for $\mu = m/2$,
increasing to $\approx 2.1$ at $\mu = 2m$.  The 1PI NNLO-NNLL $K$ factors,
$K^{(2)}$, the dashed curves, are flatter but  $K^{(2)}$ is larger than 
$K_0^{(1)}$ when $\mu = m/2$.  
However, $K^{(2)}$ is nearly independent of scale
and energy.  Including the subleading terms increases the scale dependence of
$K_{\rm sub}^{(2)}$ slightly but $K_{\rm sub}^{(2)} < K^{(2)}$ for all
energies, $K_{\rm sub}^{(2)} \approx 1.1$ for $\mu = m/2$ and $\approx 1.4$ for
$\mu = 2m$.  Note that $K^{(2)}$ and $K_{\rm sub}^{(2)}$ are both nearly
independent of $\sqrt{S}$ except for $\sqrt{S} < 20$ GeV.  The small $q
\overline q$ NNLO corrections reduce the increase of the $K$ factor at low
$\sqrt{S}$.  

\begin{figure}[htpb] 
\setlength{\epsfxsize=0.6\textwidth}
\setlength{\epsfysize=0.5\textheight}
\centerline{\epsffile{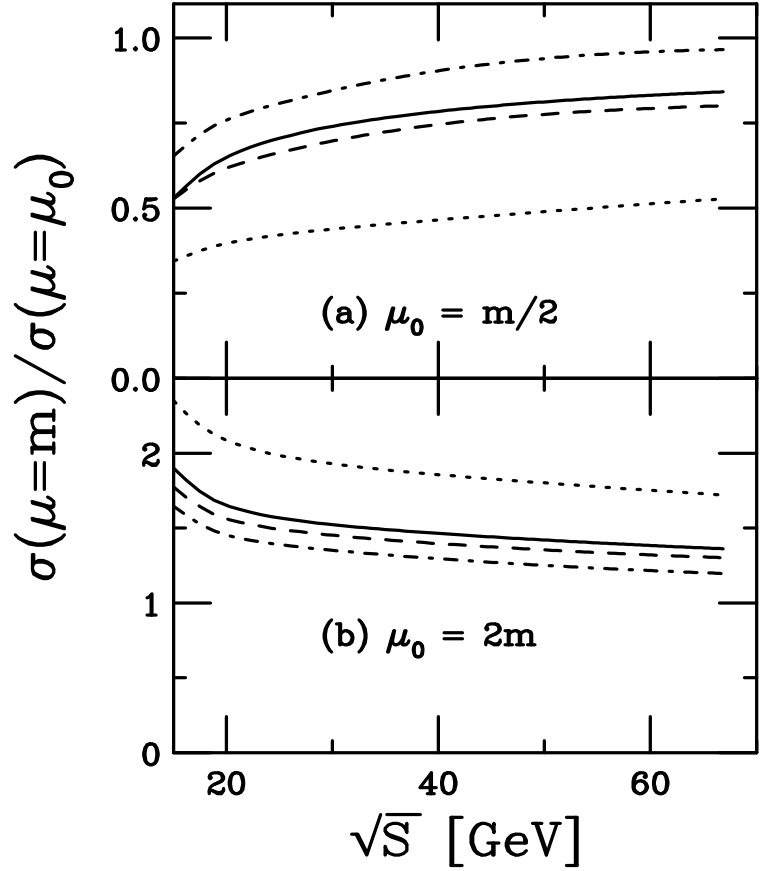}}
\caption[]{The scale dependence of $b{\bar b}$
production in $\pi^- p$ collisions with $m=4.75$ GeV. We give the ratios
$\sigma(\mu = m)/\sigma(\mu = \mu_0)$ for the LO (dotted),
NLO (solid), 1PI NNLO-NNLL (dashed) and 1PI NNLO-NNNLL$+\zeta$ (dot-dashed) 
cross sections.  The results with $\mu_0 = m/2$ are given in (a) while the
results with $\mu_0 = 2m$ are given in (b).
}
\label{f3} 
\end{figure}

Another measure of convergence is the scale dependence.  Going to higher orders
should make the result increasingly independent of scale.  We found this to be
true for $t \overline t$ production at the Tevatron \cite{KVtop}.  Here we
show the scale dependence as a function of energy for the ratio $\sigma(\mu
=m)/\sigma(\mu = \mu_0)$ where $\mu_0 = m/2$ in the upper plot and $2m$ in the
lower.  Since $\sigma(\mu = m) < \sigma(\mu = m/2)$, the upper ratio is less
than unity.  Thus an increasing independence of scale would make the ratio
approach unity from below as the cross section is calculated to higher orders.
The scale dependence is indeed decreasing for the 1PI NNLO-NNNLL$+\zeta$ cross
section over all energies.  This is a significant improvement over the
NNLO-NNLL result alone, further from unity than the NLO ratio.  Improvement is
also seen for $\mu_0 = 2m$ where the ratio should approach unity from above.

\subsection{$b$-quark production in $pp$ interactions}

We now turn to $pp$ production of bottom quarks.
So far, three experiments have reported the $b \overline b$ total cross section
in proton-induced fixed-target interactions \cite{e789,e771,herab1}, all at 
similar energies.  The cross sections are all based on relatively small event
samples.  There are two measurements at 800 GeV, the E789 result, 
using a gold target \cite{e789}, and the E771 result, using a silicon target
\cite{e771}.  There is a rather large difference between the two reported
cross sections.  The E771 measurement is in agreement with the HERA-B
measurement at 920 GeV, taken on carbon and titanium targets \cite{herab1}.
All three experiments use the $J/\psi$ decay channel.  

\begin{figure}[htpb] 
\setlength{\epsfxsize=0.6\textwidth}
\setlength{\epsfysize=0.6\textheight}
\centerline{\epsffile{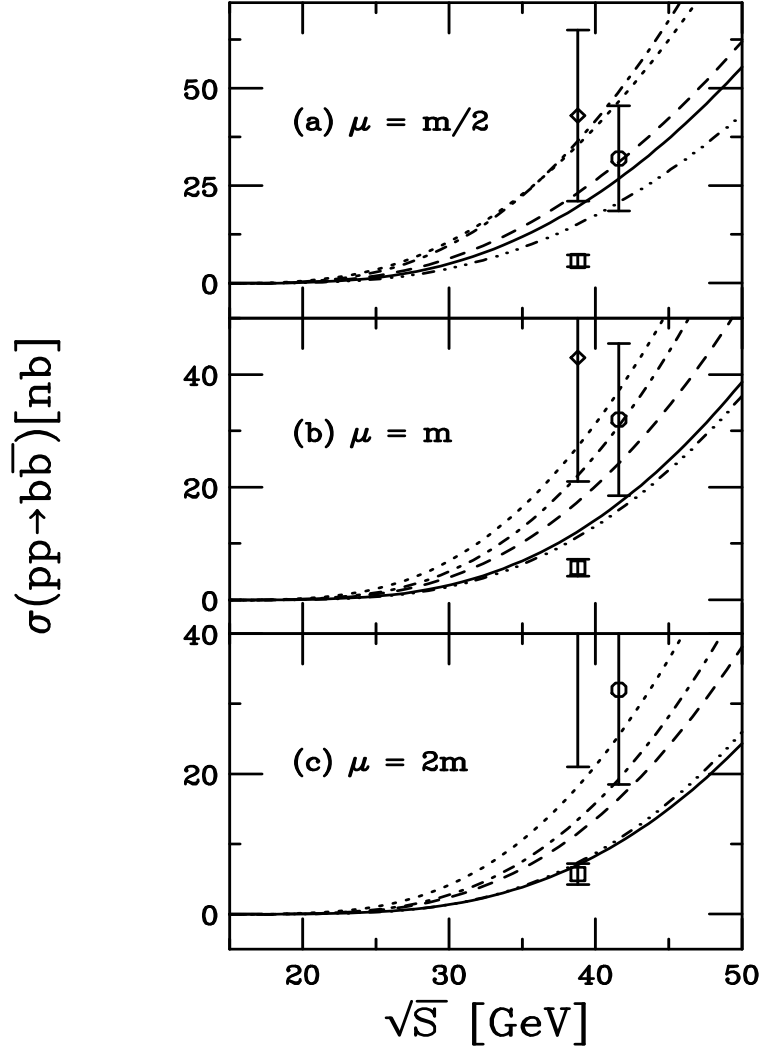}}
\caption[]{The energy dependence of $b{\bar b}$
production in $p p$ collisions with (a) $\mu = m/2$, (b) $\mu =m$ and (c)
$\mu = 2m$, calculated with the GRV98 HO proton densities. 
We show the NLO (solid) and 1PI
NNLO-NNLL (dot-dashed) results at $m = 4.75$ GeV.  The 1PI NNLO-NNNLL$+\zeta$
results are shown for $m = 4.75$ (dashed), 4.5 (dotted) and 5
(dot-dot-dot-dashed) GeV.
}
\label{f4} 
\end{figure}

\begin{figure}[htpb] 
\setlength{\epsfxsize=0.6\textwidth}
\setlength{\epsfysize=0.6\textheight}
\centerline{\epsffile{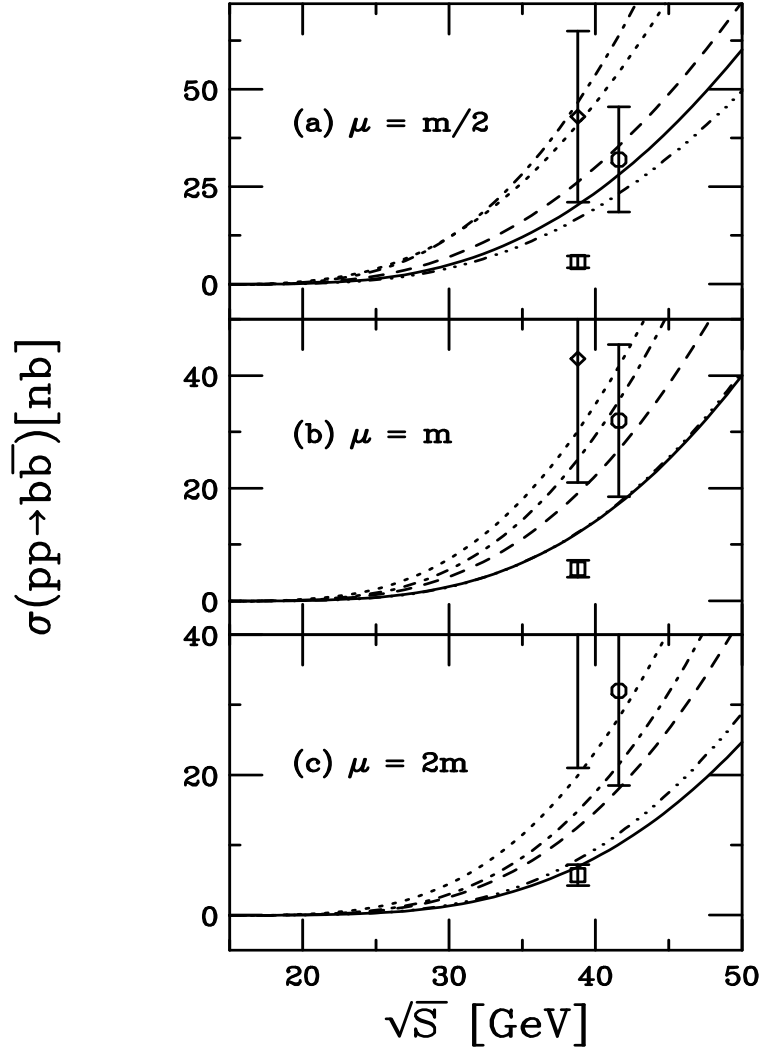}}
\caption[]{The energy dependence of $b{\bar b}$
production in $p p$ collisions with (a) $\mu = m/2$, (b) $\mu =m$ and (c)
$\mu = 2m$, calculated with the MRST2002 NNLO proton densities. 
We show the NLO (solid) and 1PI
NNLO-NNLL (dot-dashed) results at $m = 4.75$ GeV.  The 1PI NNLO-NNNLL$+\zeta$
results are shown for $m = 4.75$ (dashed), 4.5 (dotted) and 5
(dot-dot-dot-dashed) GeV.
}
\label{f5} 
\end{figure}

The three data points
are compared to our calculations with the GRV98 HO proton
parton densities in Fig.~\ref{f4} 
and with the MRST2002 NNLO parton densities in Fig.~\ref{f5}.
The results are again shown for the 
exact NLO, 1PI NNLO-NNLL and 1PI NNLO-NNNLL$+\zeta$ cross sections
at $m=4.75$ GeV as well as the 1PI NNLO-NNNLL$+\zeta$ cross sections at $m=4.5$
and 5 GeV.  

The $gg$ channel dominates $b \overline b$ production in $pp$ collisions over
the entire energy region shown.  Note that the $pp \rightarrow b \overline b$
cross section is considerably smaller than the $\pi^- p \rightarrow b \overline
b$ cross section in the near threshold region.  The $\pi^- p$ cross section is
larger because of the high valence quark-valence antiquark luminosity, absent
in $pp$ collisions.  At higher energies, far above threshold, the cross
sections in both processes become more similar.  Due to the difference
in the pion and proton gluon distributions, however, they do not become
equal. 

The trends in the calculated cross sections are similar to those of
Fig.~\ref{f1}.  Typically, the 1PI NNLO
results with the MRST2002 NNLO parton densities
are somewhat larger than those with the GRV98 HO parton densities although the
exact NLO results are quite similar.  The difference is due to the somewhat
larger value of $\Lambda_4$ for the MRST2002 NNLO densities.  The HERA-B
collaboration \cite{herab1} reported that with $m = 4.75$ GeV, $\mu = m$ and
the CTEQ5M densities \cite{cteq5}, 
the average of the 1PI and PIM NNLO-NNLL cross sections in Ref. \cite{KLMV}
was in good agreement with their data.  Since $\Lambda_4$ for the CTEQ5M
densities is larger than the corresponding values for either the GRV98 or
MRST2002 densities, the NNLO contribution is larger for the CTEQ5M densities.
The 1PI NNLO-NNNLL$+\zeta$ result, similar to the 1PI and PIM average, although
slightly lower than the mean reported HERA-B cross section in Figs.~\ref{f4} 
and \ref{f5} for the same values of $m$ and $\mu$, is still in good
agreement. 

\begin{figure}[htpb] 
\setlength{\epsfxsize=0.95\textwidth}
\setlength{\epsfysize=0.6\textheight}
\centerline{\epsffile{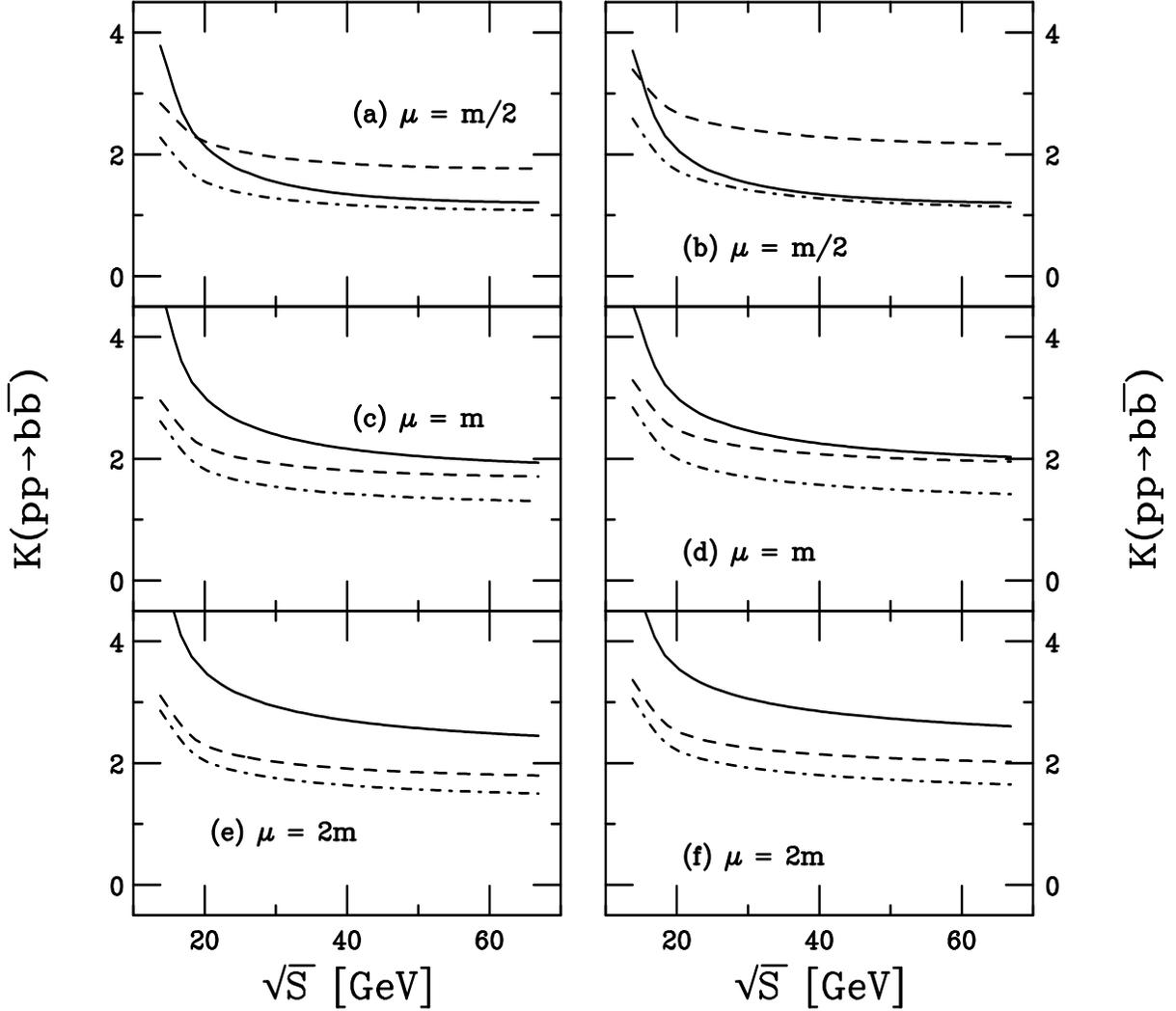}}
\caption[]{The $K$-factors for $b{\bar b}$
production in $p p$ collisions with $m=4.75$ GeV. 
We present $K_0^{(1)}$ (solid), $K^{(2)}$ (dashed) and $K_{\rm sub}^{(2)}$ 
(dot-dashed) results 
for (a) and (b), $\mu=m/2$, (c) and (d), $\mu = m$ and, (e) and (f), 
$\mu = 2m$.  Results with the GRV98 HO proton densities are shown on the
left-hand side while results with the MRST2002 NNLO proton densities are
shown on the right-hand side.
}
\label{f6} 
\end{figure}

The NNLO-NNNLL$+\zeta$ $b \overline b$ cross section at $\sqrt{S} = 41.6$ GeV 
with the MRST2002 NNLO parton densities is
\begin{eqnarray}
\sigma_{{\rm NNLO-NNNLL}+\zeta}^{\rm MRST2002 NNLO} = 28 \pm 9
\begin{array}{c} +15 \\ -10 \end{array} \, \, {\rm nb} \, \, .
\end{eqnarray}
The central value represents the result for $m = \mu = 4.75$ GeV, the first
uncertainty is due to the scale variation and the second is due to 
the variation
in $m$.  The corresponding cross section for the GRV98 densities is somewhat
smaller than the MRST result,
\begin{eqnarray}
\sigma_{{\rm NNLO-NNNLL}+\zeta}^{\rm GRV98} = 25 \begin{array}{c} +7 \\ -8
\end{array} \begin{array}{c} +13 \\ -9 \end{array} \, \, {\rm nb} \, \, .
\end{eqnarray}
The uncertainties are also reduced for this density due to the lower value of
$\Lambda_4$ associated with this set.  The NLO cross sections are, on the other
hand, essentially identical for the two sets,
\begin{eqnarray}
\sigma_{\rm NLO} = 17 \begin{array} {c} +12 \\ -7 \end{array}
\begin{array}{c} +10 \\ -6 \end{array} \, \, {\rm nb} \, \, .
\end{eqnarray}

The $K$ factors are shown for both sets of parton densities in Fig.~\ref{f6}.
Recall that when the MRST2002 NNLO densities are used, all $K$ factors are
computed with these densities, as are the cross sections at each order shown in
Fig.~\ref{f5}.  All the $K$ factors computed with both densities tend
to be larger for $pp$ than $\pi^- p$ collisions, especially at lower
$\sqrt{S}$.  They are also stronger functions of $\sqrt{S}$ than those shown in
Fig.~\ref{f2}.  The value of $K_0^{(1)}$ 
is larger than $K^{(2)}$ except for $\mu
= m/2$.  Note that $K_{\rm sub}^{(2)}$ is less than $K_0^{(1)}$ 
and $K^{(2)}$ for
all scales.  We see that $K^{(2)}$ is nearly scale independent but has a value
of $\approx 2$, still rather large.  Including subleading terms gives
$K_{\rm sub}^{(2)} \approx 1$ for $\sqrt{S} \geq 20$ GeV and $\mu = m/2$, 
increasing to $\approx 1.6$ for $\mu = 2m$.  The values of $K$ are nearly
independent of the parton densities and bottom quark mass in all cases.

\begin{figure}[htpb] 
\setlength{\epsfxsize=1.0\textwidth}
\setlength{\epsfysize=0.5\textheight}
\centerline{\epsffile{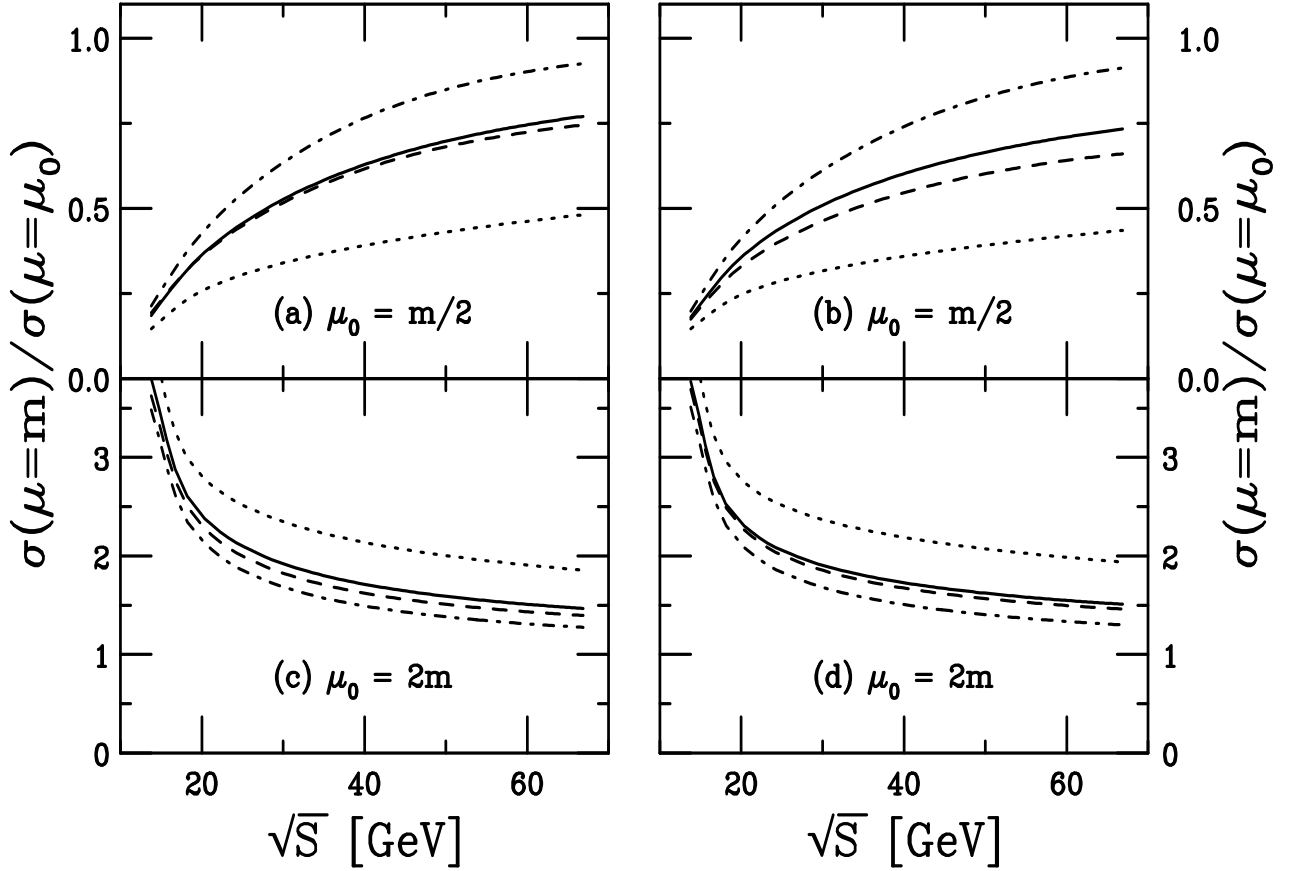}}
\caption[]{The scale dependence of $b{\bar b}$
production in $p p$ collisions with $m=4.75$ GeV. We give the ratios
$\sigma(\mu = m)/\sigma(\mu = \mu_0)$ for the LO (dotted),
NLO (solid), 1PI NNLO-NNLL (dashed) and 1PI NNLO-NNNLL$+\zeta$ (dot-dashed) 
cross sections.  The results with $\mu_0 = m/2$ are given in (a) and (b) 
while the results with $\mu_0 = 2m$ are given in (c) and (d).  Results with 
the GRV98 HO proton distributions are shown on the
left-hand side while results with the MRST2002 NNLO proton distributions are
shown on the right-hand side.
}
\label{f7} 
\end{figure}

Figure~\ref{f7} shows the scale dependence of $b \overline b$ production in
$pp$ collisions for both sets of parton densities.  We again find a 
reduction of the scale dependence for the 1PI NNLO-NNNLL$+\zeta$ results over
all energies.  Improvement in the approach of the ratio $\sigma(\mu =
m)/\sigma(\mu = \mu_0)$ toward unity is seen relative to
the 1PI NNLO-NNLL cross section ratio when $\mu_0 = m/2$.  
The latter ratio is smaller than the NLO ratio, indicating
stronger scale dependence for the 1PI NNLO-NNLL cross section than that of the
exact NLO result.  Clear improvement is also seen for $\mu_0 = 2m$.

\begin{figure}[htpb] 
\setlength{\epsfxsize=0.8\textwidth}
\setlength{\epsfysize=0.5\textheight}
\centerline{\epsffile{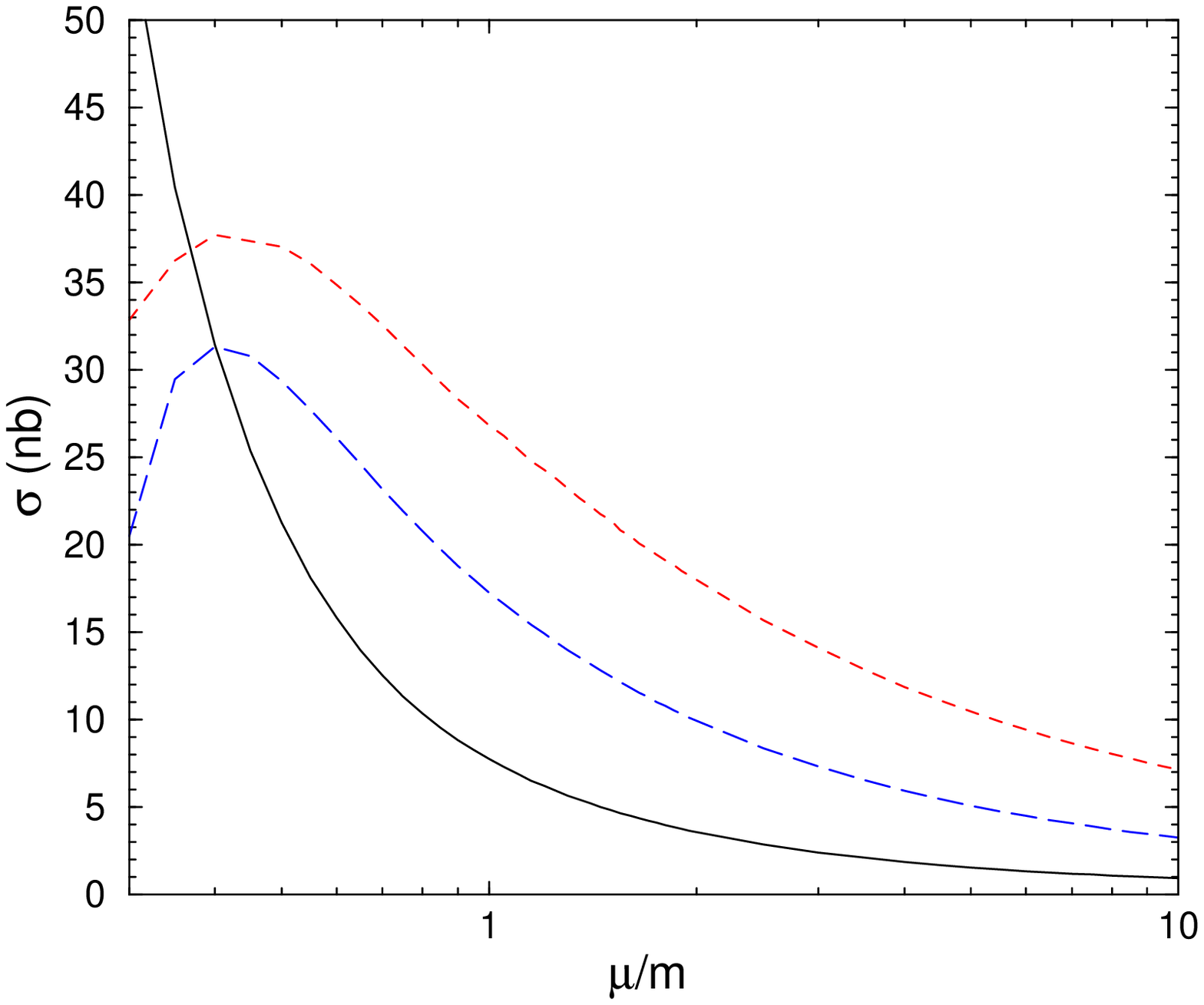}}
\caption[]{The bottom quark cross section as a function of $\mu/m$ in $pp$
collisions with $\sqrt{S}=41.6$ GeV, $m=4.75$ GeV and the MRST2002 NNLO
densities.  The Born (solid), NLO (long-dashed), 
and NNLO-NNNLL+$\zeta$ (dashed) results are shown.
}
\label{botmu} 
\end{figure}

In Fig.~\ref{botmu}, we plot the scale dependence
for $0.3 < \mu/m < 10$ with $\sqrt{S}=41.6$ GeV and
$m=4.75$ GeV. We show results for the Born, NLO, and
NNLO-NNNLL+$\zeta$ cross sections.  The scale dependence decreases with
increasing order of the cross section.  There is no plateau in the Born cross
section while the exact NLO and the NNLO-NNNLL$+\zeta$ cross sections do show
a peak in $\mu/m$.  The plateau at $\mu/m \approx 0.4$ is broader for the 
NNLO-NNNLL$+\zeta$ cross section and the overall scale dependence is reduced
relative to the exact NLO cross section.

\begin{figure}[htpb] 
\epsfxsize=0.5\textwidth\epsffile{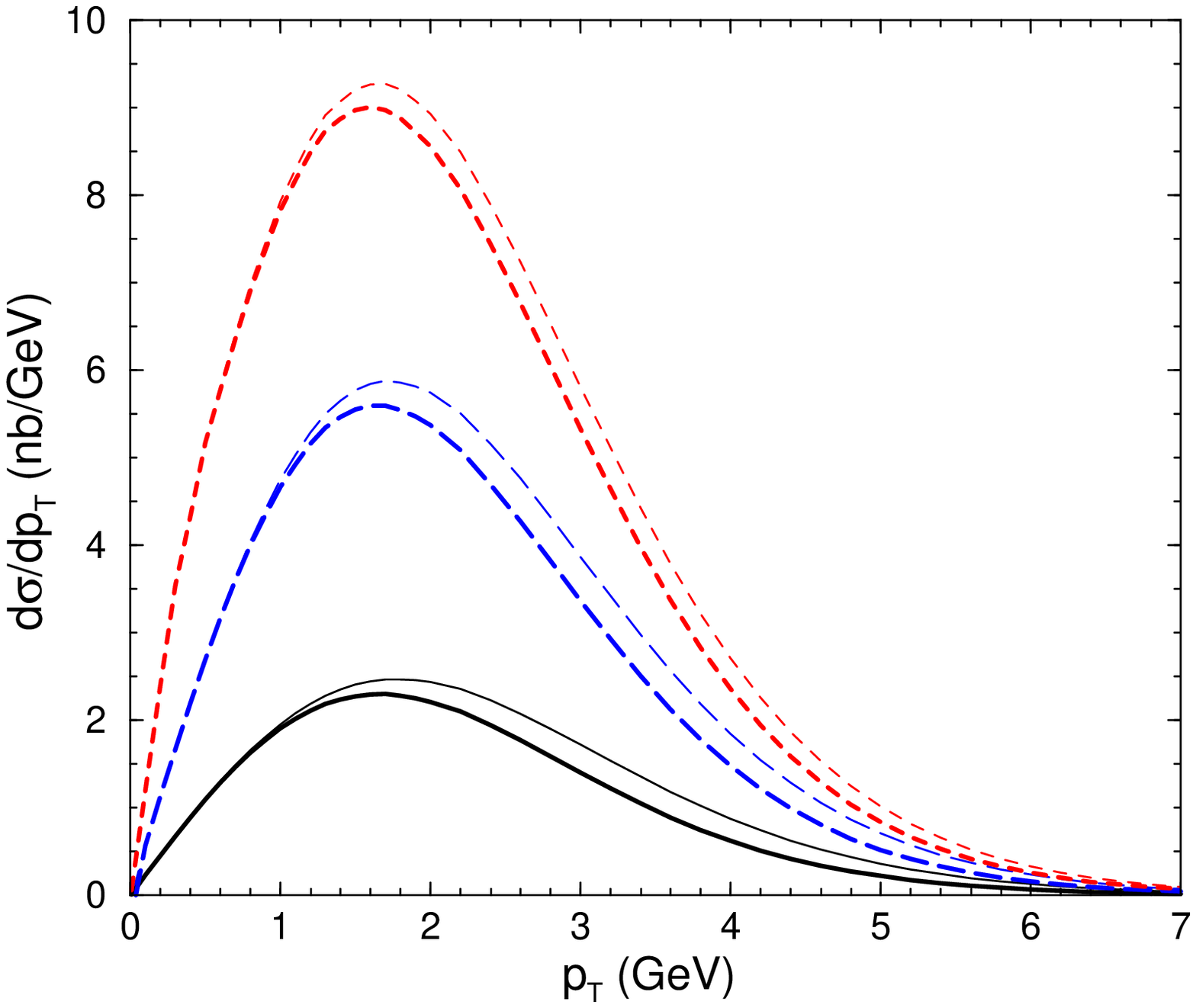}
\epsfxsize=0.5\textwidth\epsffile{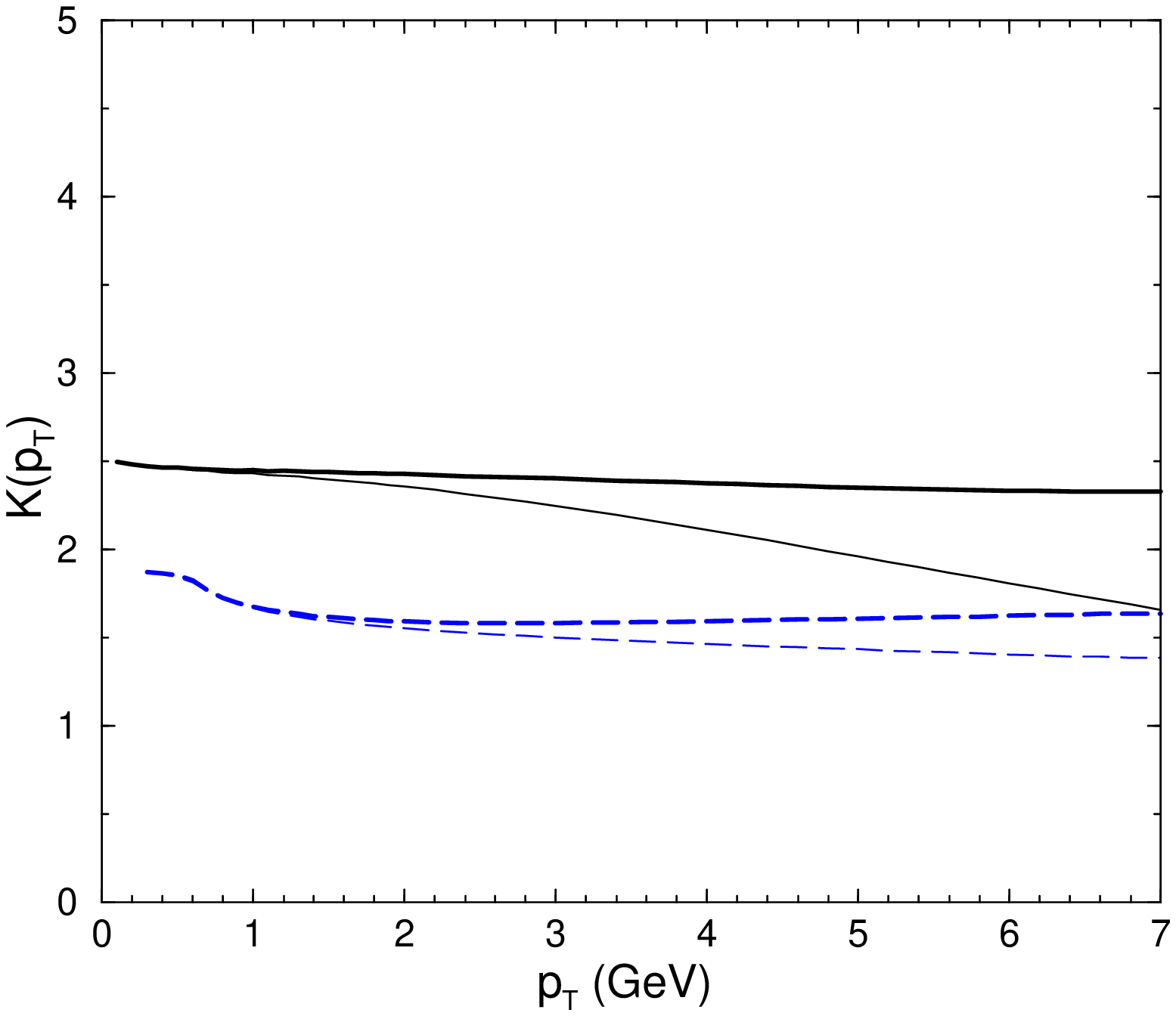}
\caption[]{The bottom quark transverse momentum distributions at $\sqrt{S} =
41.6$ GeV, $m = 4.75$ GeV and the MRST2002 NNLO parton densities.  
On the left-hand side we show the Born (solid), NLO (long-dashed), 
and NNLO-NNNLL+$\zeta$ (dashed) results while on the right-hand side we
present $K_0^{(1)}$ (solid) and $K_{\rm sub}^{(2)}$ (dashed).  The bold 
lines are calculated
with $\mu = m_T$ while the thin lines are with $\mu = m$.
}
\label{botpt} 
\end{figure}

Figure \ref{botpt} shows the $b$-quark transverse momentum distribution
at HERA-B with $\sqrt{S}=41.6$ GeV and  $m=4.75$ GeV.
The Born, NLO, and NNLO-NNNLL+$\zeta$
results are shown on the left-hand side. 
On the right-hand side we plot $K_0^{(1)}$ and $K_{\rm sub}^{(2)}$.
We provide results with two different scales,
$\mu=m$ and $\mu=m_T\equiv\sqrt{p_T^2+m^2}$.
There is some difference in the results with the two scales
at larger values of $p_T$.   The distributions with the fixed scale, $\mu = m$,
are somewhat higher than those with $\mu = m_T$.  Increasing $p_T$ and thus
$m_T$ decreases $\alpha_s$, reducing $d\sigma/dp_T$ at higher $p_T$ relative to
the fixed scale choice.  Although evolution increases the parton densities at
low $x$, at the higher momentum fractions relevant here, evolution decreases
the densities at higher scales.  Both the running of $\alpha_s$ with scale and
the evolution of the parton densities work to decrease the cross sections with
the running scale relative to the fixed scale at higher $p_T$.  With both 
choices, we see an enhancement of the bottom quark transverse momentum
distribution similar to that of the total cross sections 
when the threshold corrections are added.  
The shapes of the distributions, however, are very similar \cite{NKtop,NKJS}.

The $K$ factors are, on average,
similar to those shown in Fig.~\ref{f6} at the same energy.  
When $\mu=m$, both $K$ factors
tend to decrease with $p_T$ over the range shown.  As $p_T$ increases, the
parton momentum fractions, $x$, probed also increases reducing both the
available phase space and the $gg$ contribution to the cross section.  These
two effects work together to reduce the $K$ factors at larger $p_T$
\cite{rvzphys}.   When $\mu = m_T$, both $K$ factors are essentially constant,
as also seen in Ref.~\cite{rvzphys}.

\mysection{Charm quark production}

We now turn to charm quark production.  There is much more data on charm than
bottom production.  However, the early data are all rather low statistics and
have not been updated to include the latest branching ratios.  In our previous
report on charm production to NNLO-NNLL \cite{KLMV_cc}, we incorporated
all the data as published, making assumptions only on  how to extrapolate from
the measured $D$ meson cross sections to the charm quark cross sections.
In this paper we choose only to compare our results to the $\pi^- p$ and $pp$
data tabulated in Refs.~\cite{mlm,frix}.  These data are the most recent and
incorporate the newest measurements of branching ratios.  The $\pi^- p$ data in
the threshold region, $\sqrt{S} \leq 30$ GeV,
in order of increasing $\sqrt{S}$ are found in
Refs.~\cite{Barlag88,Alves96,Barlag91,Adamovich96,Aguilar85a} while the $pp$
data in the same energy region are given in
Refs.~\cite{Barlag88,Alves96,Aguilar88}. 

How the $c \overline c$ pairs hadronize is a 
particularly important question for energies near threshold where some channels
may be energetically disfavored.
We follow Ref.~\cite{mlm} and assume that $\sigma(D_s)/\sigma(D^0 + 
D^+) \simeq 0.2$ and $\sigma(\Lambda_c)/\sigma(D^0 + D^+) \simeq 0.3$,
independent of energy, so that
the total $c \overline c$ cross section is obtained from 
$\approx 1.5 \, \sigma(D \overline D)$.  This assumption could have a strong 
energy dependence near threshold.  Thus
as many charm hadrons (mesons and baryons) as possible should be measured to
better understand fragmentation and hadronization.  Finally,
some of the data are taken on nuclear targets and then extrapolated to $\pi^-
p$ and $pp$ assuming a linear $A$ dependence \cite{alv,Leitch94}.

Recent comparisons of the full $pp \rightarrow c \overline c$ data set with
exact NLO cross sections were made to determine the best mass and scale choices
for extrapolation to higher energies \cite{RVhpc}.  Rough
agreement with the data up to the top ISR energy, $\sqrt{S} = 63$ GeV, was
found for $m = 1.2$ GeV and $\mu = 2m$ for the MRST densities and $m = 1.3$ GeV
with $\mu = m$ for the GRV98 densities \cite{RVhpc}.  These values of $m$
are rather small compared to the typical value of 1.5 GeV.  Thus, as in our
previous paper \cite{KLMV_cc}, we calculate the NLO, 1PI NNLO-NNLL and, in
addition here, the 1PI NNLO-NNNLL$+\zeta$ cross sections using $m = 1.2$, 1.5
and 1.8 GeV as well as $\mu = m$ and $2m$.  We can then test whether the 
NNLO+NNNLL$+\zeta$ cross sections might favor a 
higher charm quark mass.  Our charm
calculations also employ the GRV98 HO and MRST2002 NNLO proton parton
densities. 

\subsection{Charm production in $\pi^- p$ interactions}

\begin{figure}[htpb] 
\setlength{\epsfxsize=0.95\textwidth}
\setlength{\epsfysize=0.6\textheight}
\centerline{\epsffile{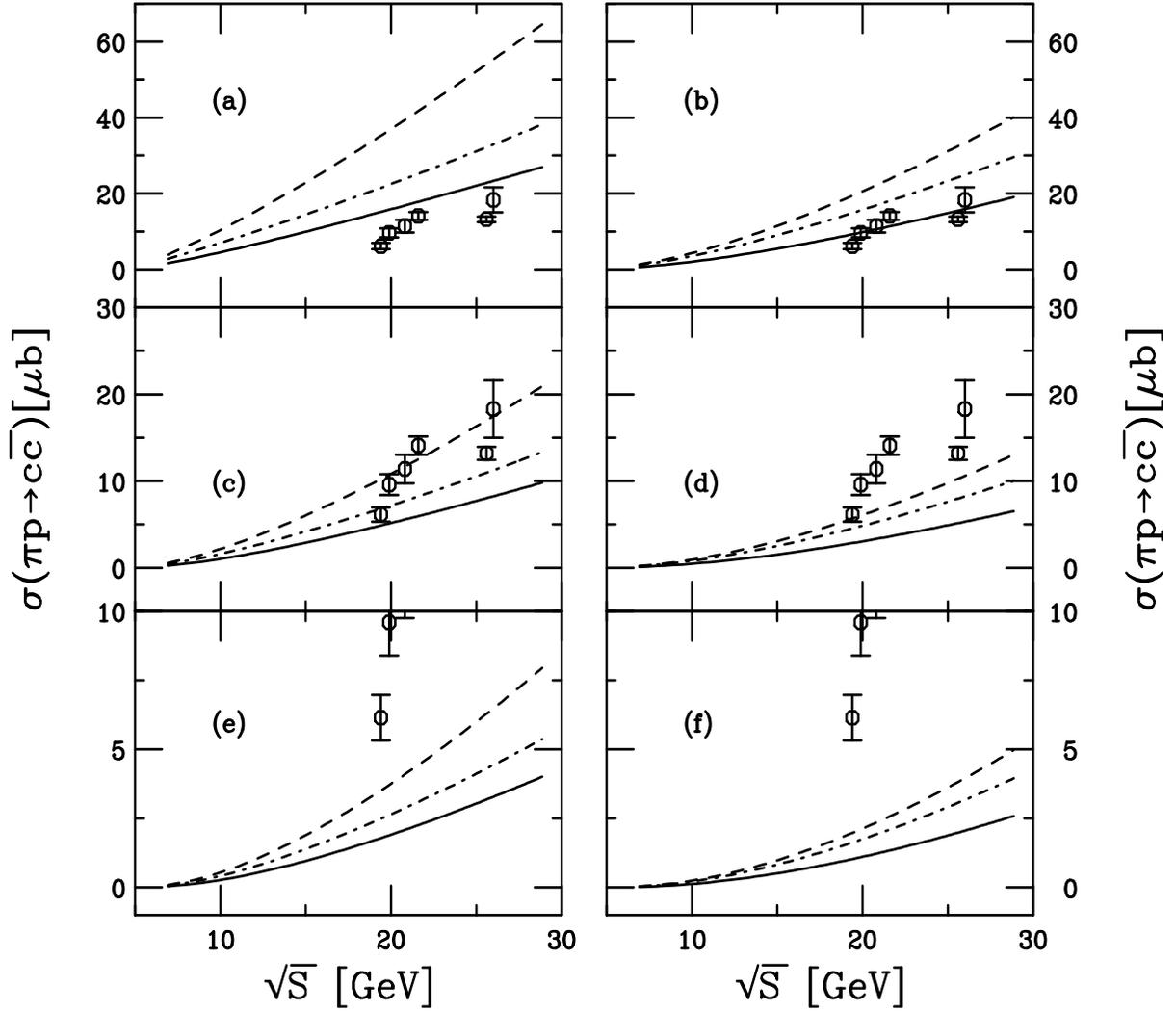}}
\caption[]{The energy dependence of $c{\bar c}$
production in $\pi^- p$ collisions with, (a) and (b), $m = 1.2$, (c) and (d),
$m = 1.5,$ and, (e) and (f), $m = 1.8$ GeV, calculated with the GRV98 HO 
proton densities and the GRS pion densities.  We show the NLO (solid),
1PI NNLO-NNLL (dashed) and 1PI NNLO-NNNLL$+\zeta$ (dot-dashed)
results.  On the left-hand side, $\mu = m$ while on the right-hand
side, $\mu = 2m$. 
}
\label{f8} 
\end{figure}

We first discuss the results from $\pi^- p$ interactions.  These data are
typically reported as the forward cross sections, $x_F > 0$.  The cross section
is extrapolated to all $x_F$ assuming that $\sigma({\rm all}\,\, x_F) = 1.6
\sigma(x_F>0)$, calculated to NLO.  The difference is not a factor of two as
in $pp$ production since the forward $x_F$ distribution is harder than the
backward distribution due to the harder large $x$ behavior of the pion gluon
distribution. 

The reaction $\pi^- p \rightarrow c \overline c$ is dominated by the $gg$
channel for $\sqrt{S} \geq 15$ GeV \cite{SVchm}.  The $gg$ and $q \overline q$
channels are only similar in magnitude very close to threshold.    The
comparison of the data with the exact NLO, 1PI NNLO-NNLL and 1PI
NNLO-NNNLL$+\zeta$ cross sections, calculated with the GRV98 HO proton parton
densities and the GRS pion parton densities, is shown in Fig.~\ref{f8}.
The exact NLO and the 1PI NNLO-NNLL results were previously shown in
Ref.~\cite{KLMV_cc}.  The new 1PI NNLO-NNNLL$+\zeta$ results, indicated by the
dot-dashed curves, are similar to the average of the NNLO-NNLL 1PI and PIM
cross sections of Ref.~\cite{KLMV_cc}.  At the higher end of the $\sqrt{S}$
range studied, the new results with the subleading logs are higher than
this average because the PIM NNLO-NNLL $gg$ contribution is large and negative
above threshold. Eventually this contribution becomes larger than the exact NLO
cross section, resulting in a negative PIM NNLO-NNLL total cross section and a
decreased 1PI and PIM average NNLO-NNLL cross section, particularly for the
lower values of $m$.
Inclusion of the subleading logs mitigates this behavior \cite{KVtop} albeit
not enough to make the PIM calculation reliable for gluon dominated processes.

There is relatively good agreement between the exact NLO calculations and the
data for $m = 1.2$ GeV and $\mu = 2m$, Fig.~\ref{f8}(b).  We can see,
however, that the data are also relatively compatible with the 1PI NNLO cross
sections, both the NNLO-NNLL and NNLO-NNNLL$+\zeta$ calculations, when $m = \mu
= 1.5$ GeV.  This result suggests that a full NNLO calculation would be more
compatible with a larger charm quark mass.

\begin{figure}[htpb] 
\setlength{\epsfxsize=0.95\textwidth}
\setlength{\epsfysize=0.6\textheight}
\centerline{\epsffile{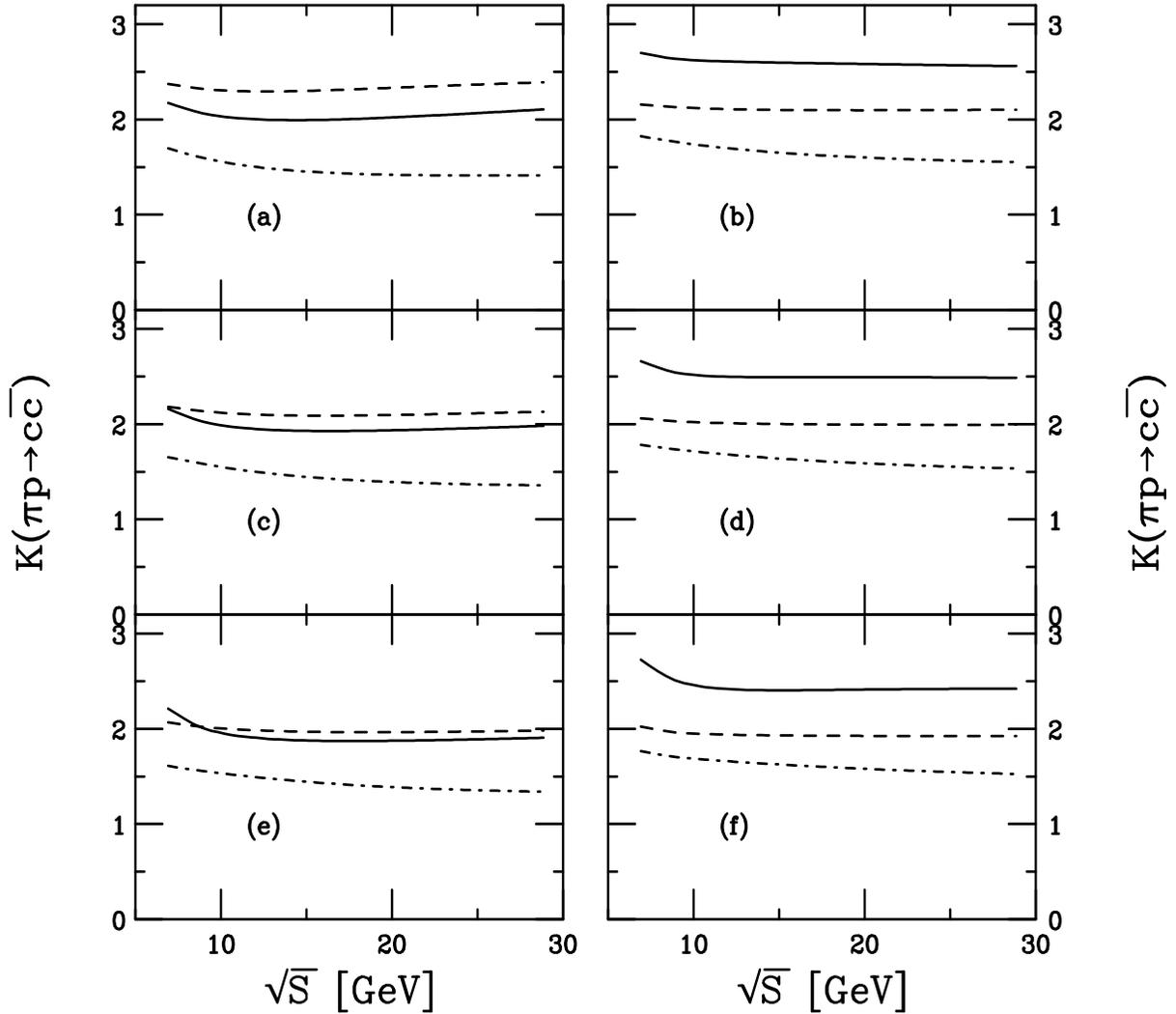}}
\caption[]{The $K$-factors for $c{\bar c}$
production in $\pi^- p$ collisions with, (a) and (b), $m = 1.2$, (c) and (d),
$m = 1.5,$ and, (e) and (f), $m = 1.8$ GeV.
We present $K_0^{(1)}$ (solid), $K^{(2)}$ (dashed) and $K_{\rm sub}^{(2)}$ 
(dot-dashed) for $\mu=m$ (left-hand side) and $\mu = 2m$ (right-hand side).
}
\label{f9} 
\end{figure}

The $\pi^- p$ charm production $K$ factors are shown in Fig.~\ref{f9}.  As
expected, the $K$ factors are all larger for charm than bottom production due
to the dominance of the $gg$ channel over all $\sqrt{S}$ as well as the
larger value of $\alpha_s$.  
There is a significant decrease in the $K$ factor when the subleading terms
are included: $K_{\rm sub}^{(2)} \approx 1.5$ whereas $K^{(2)} \approx 2$. 
All the $K$ factors decrease slightly with increasing $m$.  The mass effect
is smaller on $K_0^{(1)}$ than on $K^{(2)}$, the 1PI NNLO-NNLL $K$
factor.  Indeed the $K^{(2)}$ mass dependence is larger than either that of
$K_0^{(1)}$ or $K_{\rm sub}^{(2)}$.
The exact NLO $K$ factor, $K_0^{(1)}$, shows the strongest scale
dependence, increasing from $\approx 2$ for $\mu = m$
to $\approx 2.5$ when $\mu = 2m$, as
also seen for $b \overline b$ production.  Note also that while all the $K$
factors are relatively energy independent, $K_0^{(1)}$ exhibits the largest
$\sqrt{S}$ dependence. 

\begin{figure}[htpb] 
\setlength{\epsfxsize=0.6\textwidth}
\setlength{\epsfysize=0.6\textheight}
\centerline{\epsffile{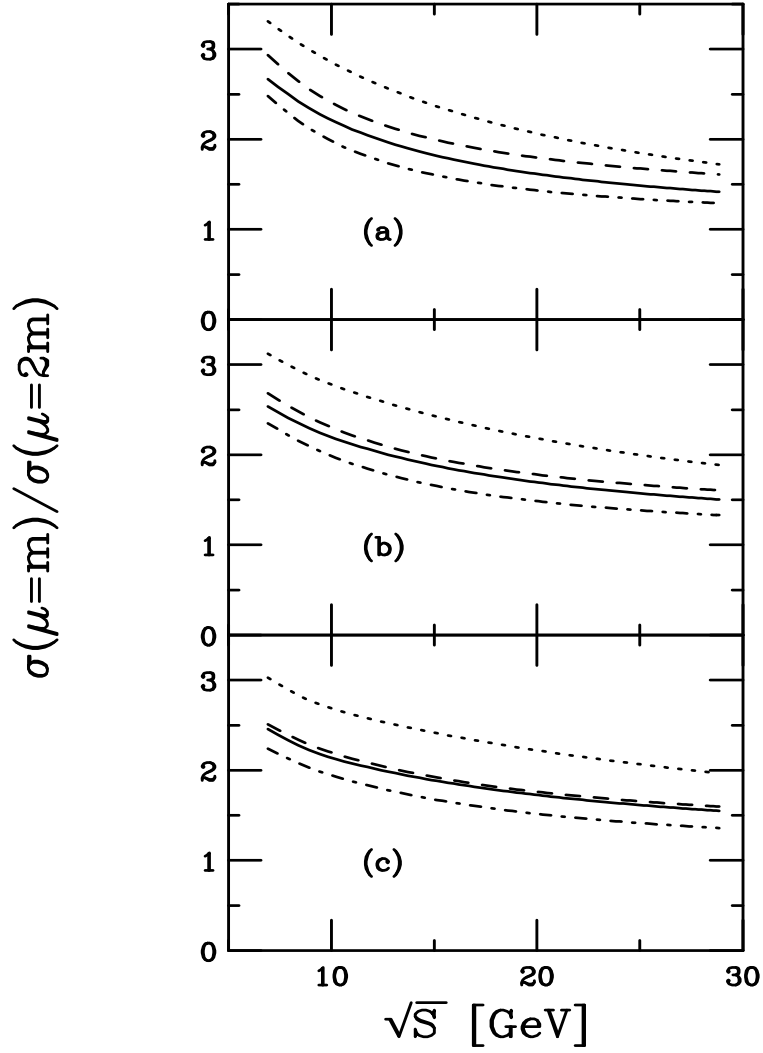}}
\caption[]{The scale dependence of $c{\bar c}$
production in $\pi^- p$ collisions with (a) $m = 1.2$, (b) $m = 1.5,$ and (c) 
$m = 1.8$ GeV.  We give the ratios
$\sigma(\mu = m)/\sigma(\mu = 2m)$ for the LO (dotted),
NLO (solid), 1PI NNLO-NNLL (dashed) and 1PI NNLO-NNNLL$+\zeta$ (dot-dashed) 
cross sections.  
}
\label{f10} 
\end{figure}

Figure~\ref{f10} demonstrates the scale dependence of charm production in
$\pi^- p$ interactions.  Since $\mu = m/2$ is less than the initial scale of
most sets of parton densities, this value of $\mu$ is essentially meaningless
for charm production.  Therefore, we plot the ratio $\sigma(\mu = m)/\sigma(\mu
= 2m)$ as a function of $\sqrt{S}$ for all three values of $m$.  The scale
dependence only weakly depends on charm mass although the dependence is
slightly stronger at low $\sqrt{S}$.  The 1PI NNLO-NNNLL$+\zeta$ ratio shows
significant improvement over both the NLO and the 1PI NNLO-NNLL results. 

\subsection{Charm production in $pp$ interactions}

\begin{figure}[htpb] 
\setlength{\epsfxsize=0.95\textwidth}
\setlength{\epsfysize=0.6\textheight}
\centerline{\epsffile{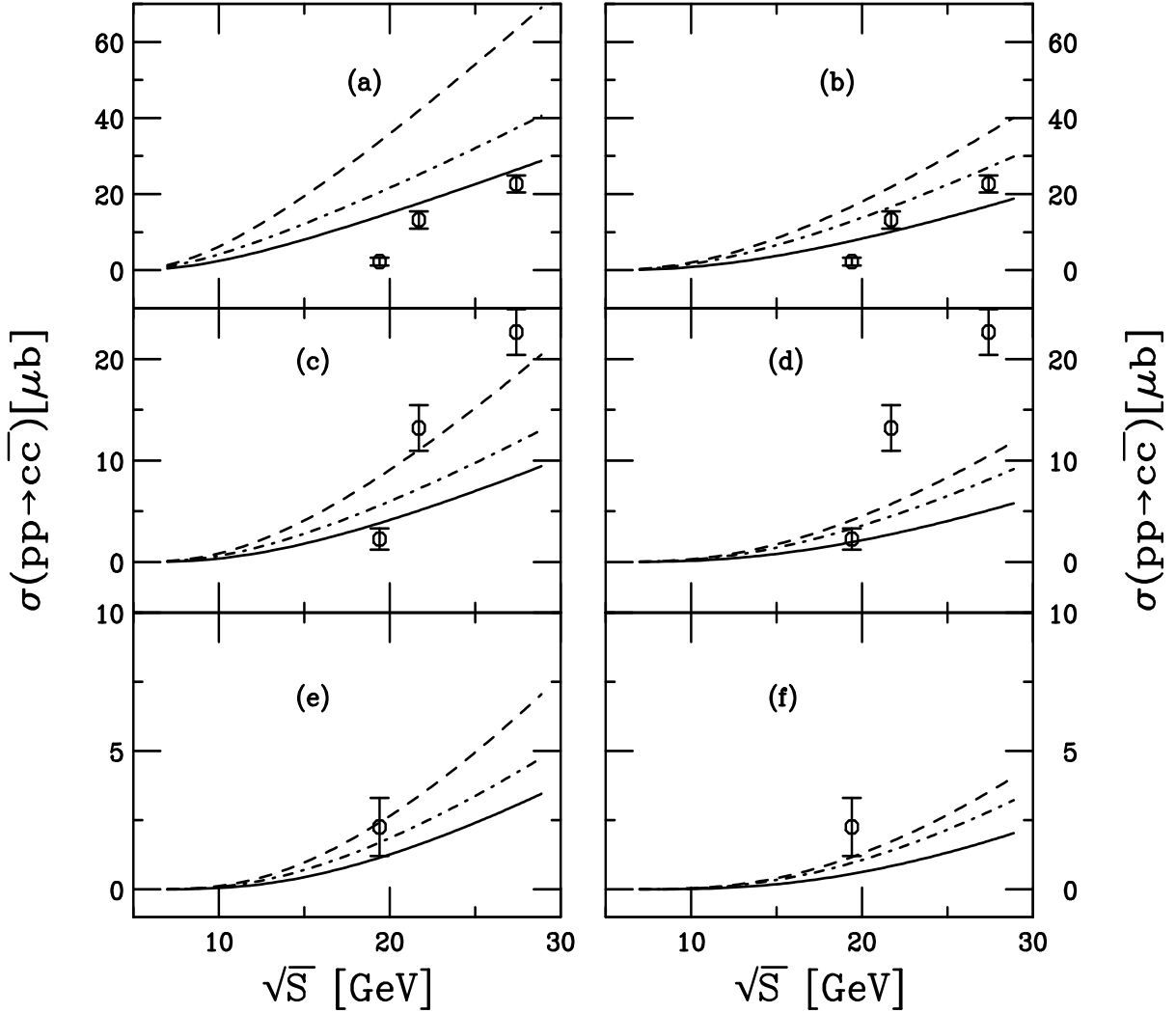}}
\caption[]{The energy dependence of $c{\bar c}$
production in $p p$ collisions with, (a) and (b), $m = 1.2$, (c) and (d),
$m = 1.5,$ and, (e) and (f), $m = 1.8$ GeV, calculated with the GRV98 HO 
proton densities.  We show the NLO (solid),
1PI NNLO-NNLL (dashed) and 1PI NNLO-NNNLL$+\zeta$ (dot-dashed)
results.  On the left-hand side, $\mu = m$ while on the right-hand
side, $\mu = 2m$. 
}
\label{f11} 
\end{figure}

\begin{figure}[htpb] 
\setlength{\epsfxsize=0.95\textwidth}
\setlength{\epsfysize=0.6\textheight}
\centerline{\epsffile{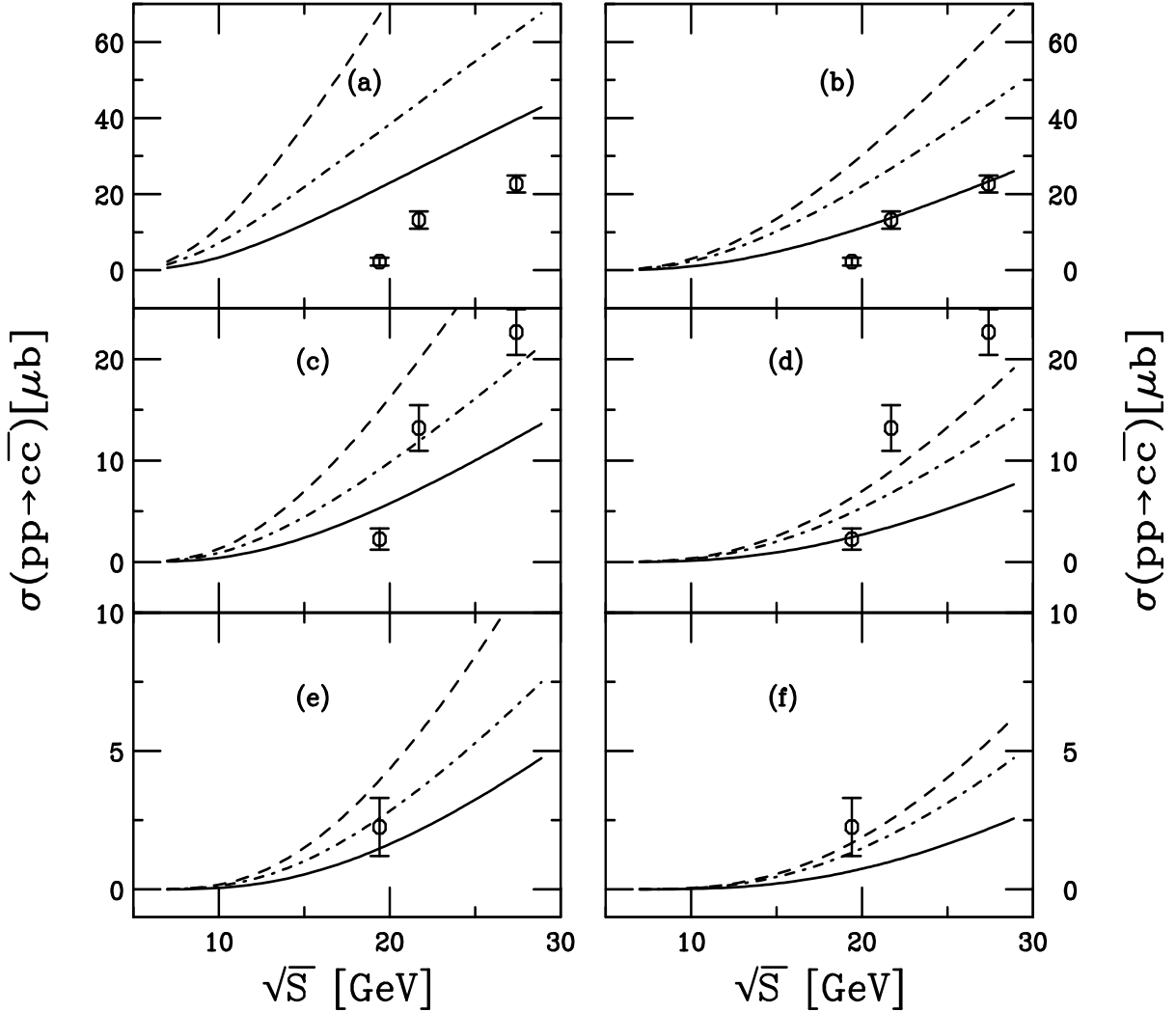}}
\caption[]{The energy dependence of $c{\bar c}$
production in $p p$ collisions with, (a) and (b), $m = 1.2$, (c) and (d),
$m = 1.5,$ and, (e) and (f), $m = 1.8$ GeV, calculated with the MRST2002 NNLO
proton densities.  We show the NLO (solid),
1PI NNLO-NNLL (dashed) and 1PI NNLO-NNNLL$+\zeta$ (dot-dashed)
results.  On the left-hand side, $\mu = m$ while on the right-hand
side, $\mu = 2m$.
}
\label{f12} 
\end{figure}

We now consider $pp \rightarrow c \overline c$ interactions.  In
Figs.~\ref{f11} and \ref{f12}, we compare the exact NLO, 1PI NNLO-NNLL and 1PI
NNLO-NNNLL$+\zeta$ cross sections calculated with the GRV98 HO and MRST2002
NNLO proton parton densities respectively.  There is a larger difference
between the results with the two parton densities than seen for $b \overline b$
production.  Since $\Lambda_3 > \Lambda_4$ the overall NNLO corrections are
thus larger for charm than bottom as well, as evident from the larger charm $K$
factors.  At NLO, the best agreement is again with $m = 1.2$ GeV and $\mu =
2m$, seen in Figs.~\ref{f11} and \ref{f12} (b).  The new result in
Fig.~\ref{f11} is the subleading NNLO-NNNLL$+\zeta$
cross section, the other calculations,
also shown in Ref.~\cite{KLMV_cc}, are repeated for comparison purposes.  The
results in Fig.~\ref{f12} are all new. 

The MRST2002 NNLO parton densities generally give larger cross sections, even
for the exact NLO result since the value of $\Lambda_3$ is larger than that of
the GRV98 HO set.  Also due the larger $\Lambda_3$, the NNLO corrections are
significantly larger, as seen in Fig.~\ref{f12}.  A similar effect was seen in
the comparison of the GRV98 HO and CTEQ5M results in Ref.~\cite{KLMV_cc}.

In both Figs.~\ref{f11} and \ref{f12}, the exact NLO cross sections calculated
with $m = 1.2$ GeV and $\mu = 2m$ are relatively compatible with the data.  
Note that in Fig.~\ref{f11}(b), the 1PI NNLO-NNNLL$+\zeta$ result is in
somewhat better agreement with the two highest energy data points than the
exact NLO.  The
1PI NNLO cross sections are in rather good agreement with the data when $m =
\mu = 1.5$ GeV is used with the MRST2002 NNLO parton densities.  Indeed, 
the 1PI NNLO-NNNLL$+\zeta$ result in
Fig.~\ref{f12}(c) agrees rather well with the two higher energy data points.
Thus, as in Ref.~\cite{KLMV_cc}, 
we can conclude that the full NNLO result can likely describe the charm
data well with $m = \mu = 1.5$ GeV whereas the lower mass is needed with an NLO
calculation alone. 

\begin{figure}[htpb] 
\setlength{\epsfxsize=0.95\textwidth}
\setlength{\epsfysize=0.6\textheight}
\centerline{\epsffile{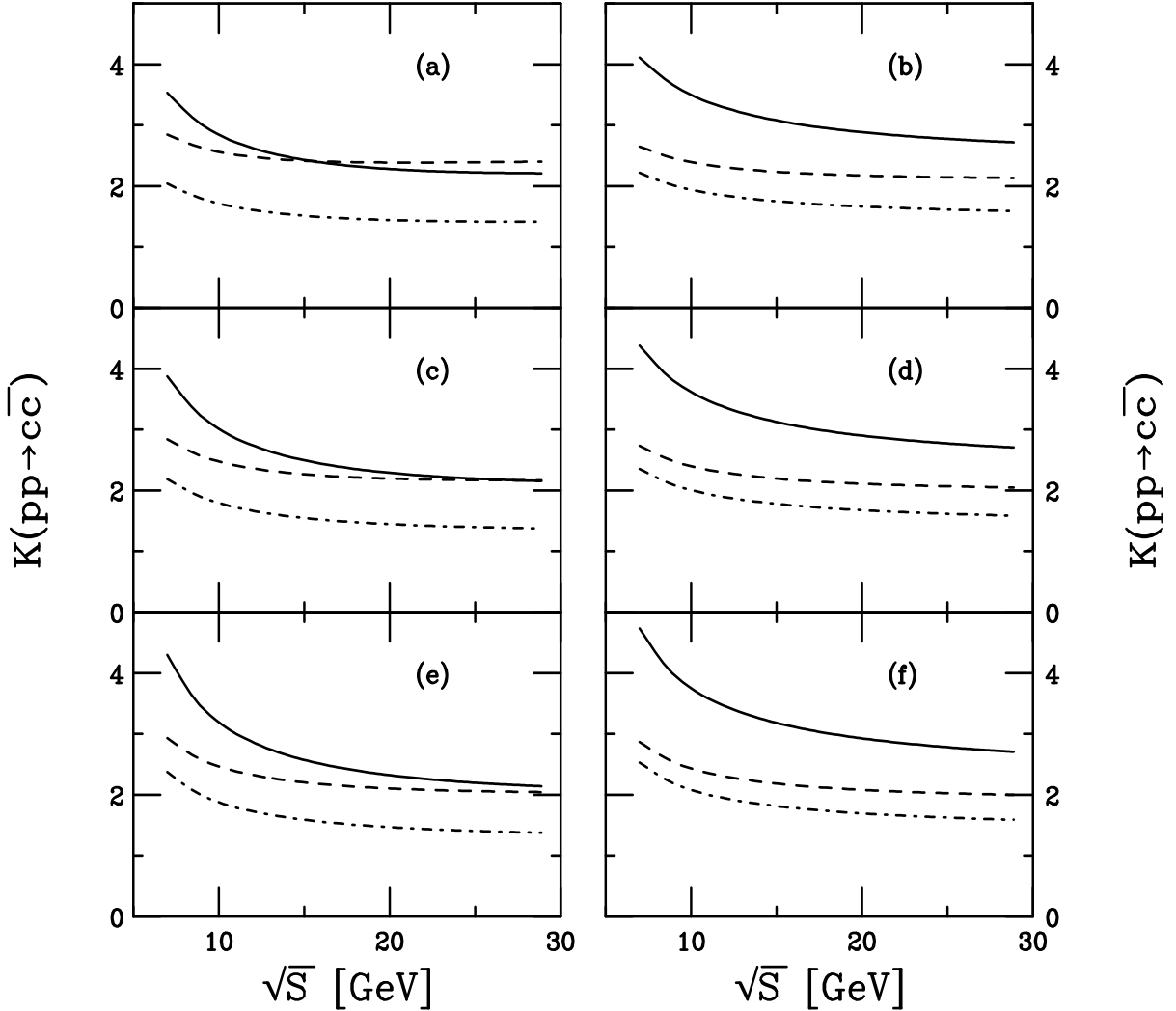}}
\caption[]{The $K$-factors for $c{\bar c}$
production in $pp$ collisions with, (a) and (b), $m = 1.2$, (c) and (d),
$m = 1.5,$ and, (e) and (f), $m = 1.8$ GeV, calculated with the GRV98 HO
parton densities.  We present $K_0^{(1)}$ (solid), $K^{(2)}$ (dashed) and 
$K_{\rm sub}^{(2)}$ (dot-dashed)
for $\mu=m$ (left-hand side) and $\mu = 2m$ (right-hand side).
}
\label{f13} 
\end{figure}

Table~\ref{cctable} gives the charm cross sections in $pp$ collisions at two
energies, $\sqrt{S} = 6.98$ GeV, the future GSI energy, and 17.3 GeV, the CERN
SPS ion energy.  We present results for both the MRST2002 NNLO and GRV98 parton
densities based on a central value of $m = \mu = 1.5$ GeV.  (The choice of
mass and scale used for our central value is for better illustration of the 
uncertainties rather than any fit to data.)  The first
uncertainty is due the the scale choice.  Since we do not calculate the result
for $\mu = m/2$ here, we show only the difference between the values of $\mu =
m$ and $2m$. The second uncertainty is that due to the charm quark mass.  The
exact NLO and the 1PI NNLO-NNNLL$+\zeta$ cross sections are shown.  Note that
the differences between the MRST2002 and GRV98 densities are larger for charm
since the difference in the values of $\Lambda_3$ between the two sets is 
larger
than that for $\Lambda_4$.  The relative increase of the NNLO-NNNLL$+\zeta$
cross section is highest at the lower energy with an increase over the NLO
result by a factor of 2.6 for the MRST densities and 2.2 for the GRV98.  
At 17.3
GeV, the increase is by a factor of 1.8 and 1.5 respectively.  Thus the
importance of the near-threshold corrections is reduced further from the charm
production threshold, as might be expected.

\begin{table}[htb]
\begin{center}
\begin{tabular}{|c|c|c|c|} \hline
\multicolumn{4}{|c|}{$\sigma$ $(\mu$b)} \\ \hline
$\sqrt{S}$ (GeV) & Order & MRST2002 NNLO & GRV98 \\ \hline
6.98 & NLO                & $0.034 \; \; - 0.027 
\begin{array}{c} +0.56 \\ \; -0.032
\end{array}$ & $0.028 \; \; - 0.022 
\begin{array}{c} +0.42 \\ \; -0.026 \end{array}$
\\ \hline 
6.98 & NNLO-NNNLL$+\zeta$ & $0.09 \; \; - 0.07 
\begin{array}{c} +1.4 \\ \; \; -0.085
\end{array}$ & $0.061 \; \; - 0.05 
\begin{array}{c} +0.9 \\ \; \; -0.057 \end{array}$
\\ \hline  \hline
17.3 & NLO                & $3.8 \; \; - 2.1 
\begin{array}{c} \; +13 \\ \; \, -2.8 \end{array}$ 
& $2.8 \; \; - 1.4 \begin{array}{c} \; \; +8.3 \\ \; -2 \end{array}$ \\ 
\hline 
17.3 & NNLO-NNNLL$+\zeta$ & $6.7 \; \; - 3.4 
\begin{array}{c} \; \; +22.5 \\ \; - 4.9
\end{array}$ & $4.1 \; \; - 1.8 
\begin{array}{c} \; \; +12.2 \\ -3 \end{array}$ \\ 
\hline 
\end{tabular}
\caption[]{The $c \overline c$ production cross sections 
in $pp$ collisions at $\sqrt{S} = 6.98$ and 17.3 GeV.
The exact NLO results and the approximate NNLO-NNNLL$+\zeta$ results,
based on $m = \mu = 1.5$ GeV, are shown.  The first uncertainty is due to the 
scale choice, the second, the charm quark mass.}
\label{cctable}
\end{center}
\end{table}

\begin{figure}[htpb] 
\setlength{\epsfxsize=0.95\textwidth}
\setlength{\epsfysize=0.6\textheight}
\centerline{\epsffile{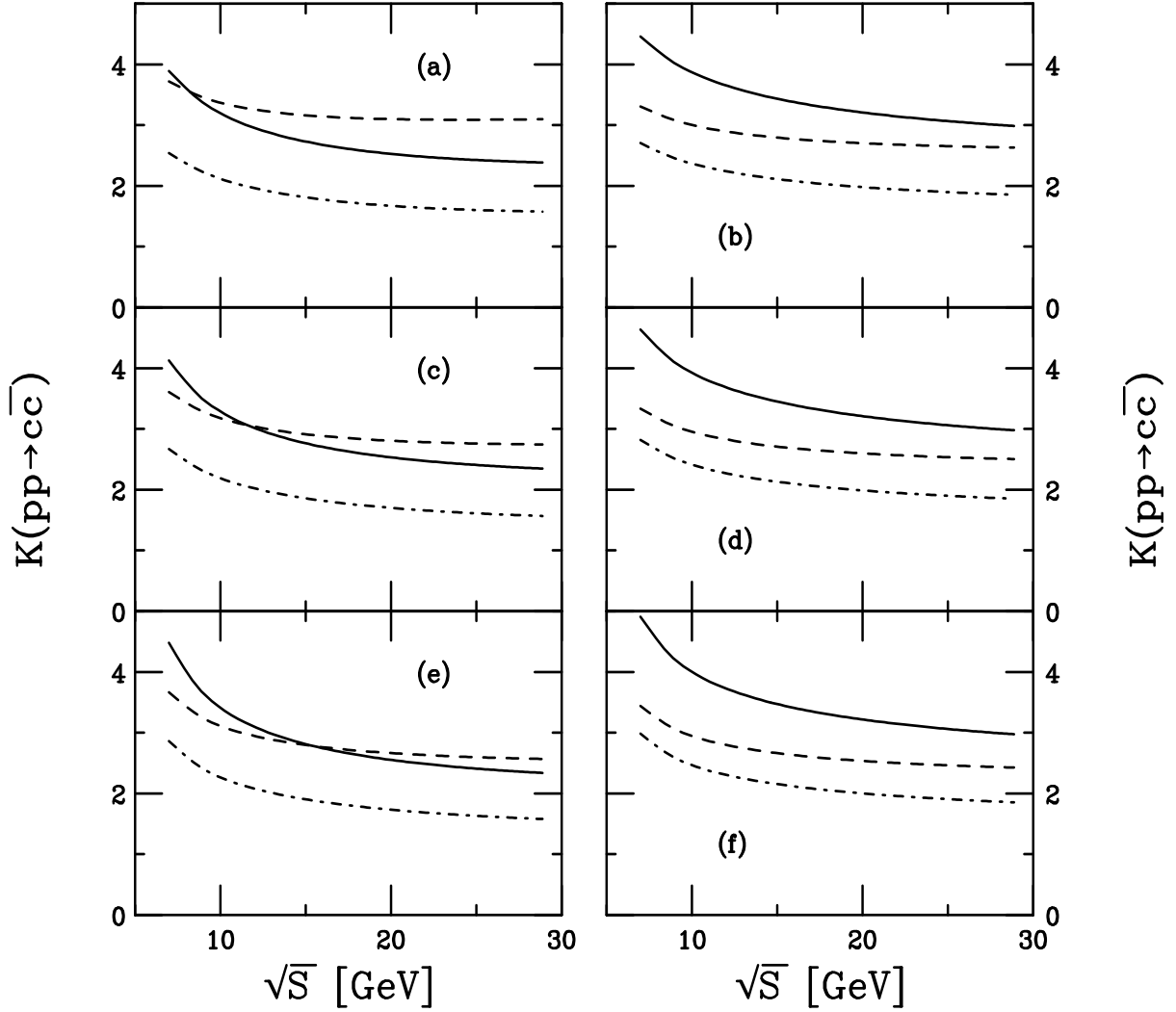}}
\caption[]{The $K$-factors for $c{\bar c}$
production in $pp$ collisions with, (a) and (b), $m = 1.2$, (c) and (d),
$m = 1.5,$ and, (e) and (f), $m = 1.8$ GeV, calculated with the MRST2002 NNLO
parton densities.  We present $K_0^{(1)}$ (solid), $K^{(2)}$ (dashed) and 
$K_{\rm sub}^{(2)}$ (dot-dashed) 
for $\mu=m$ (left-hand side) and $\mu = 2m$ (right-hand side).
}
\label{f14} 
\end{figure}

In Figs.~\ref{f13} and \ref{f14}, we compare the $pp$ $K$ factors for GRV98 HO
and MRST2002 NNLO respectively.  The same trends are seen as in $\pi^- p$
interactions although the energy dependence at low $\sqrt{S}$, particularly for
$K^{(1)}$, is stronger.  None of the $K$ factors are strong functions of mass,
scale or parton density.  The parton density dependence is thus reduced
relative to the calculations with CTEQ5M shown in Ref.~\cite{KLMV_cc}.  This is
perhaps because the CTEQ5M gluon density is higher than either the GRV98 HO or
MRST2002 NNLO gluon densities at $x \leq 0.1$ and at the low charm production
scales.  Although the initial scales of the CTEQ5 and MRST sets are not very
different, 1 GeV$^2$ for CTEQ5M and 1.25 GeV$^2$ for MRST2002, the low $x$
behavior is quite different.  Indeed at low scales, when $m = \mu = 1.2$ GeV,
the CTEQ5M gluon distribution is essentially constant as $x \rightarrow 0$ with
the tail of the distribution appearing at $x \approx 0.1$, while for the
MRST2002 NNLO set the gluon distribution becomes negative at $x < 0.001$ and
peaks at $x < 0.1$.  When $\mu$ is increased to $2m$ for the same mass, the
CTEQ5M density decreases with increasing $x$ while the MRST2002 NNLO density
peaks around $x \approx 0.001$, decreasing at lower $x$.  Although our
calculations are not sensitive to the very low $x$ region, the different 
low-$x$ behavior affects the gluon distributions at higher $x$.  The lower MRST
gluon density, combined with the lower $\Lambda_3$ relative to CTEQ5M reduces
the higher order corrections and hence the $K$ factors.  Thus $K_{\rm
sub}^{(2)}$ varies between 1.3 and 1.7 for GRV98 HO and 1.5 to 1.9 for MRST2002
NNLO when $\sqrt{S} \geq 15$ GeV, 
significant improvements over the 1PI NNLO-NNLL results, $K^{(2)} \approx
2-2.4$ for GRV98 HO and $2.5-3$ for MRST2002 NNLO and CTEQ5M.  In addition,
$K_{\rm sub}^{(2)} < K_0^{(1)}$ for all cases considered, not a feature of the
1PI NNLO-NNLL result. 

\begin{figure}[htpb] 
\setlength{\epsfxsize=0.95\textwidth}
\setlength{\epsfysize=0.6\textheight}
\centerline{\epsffile{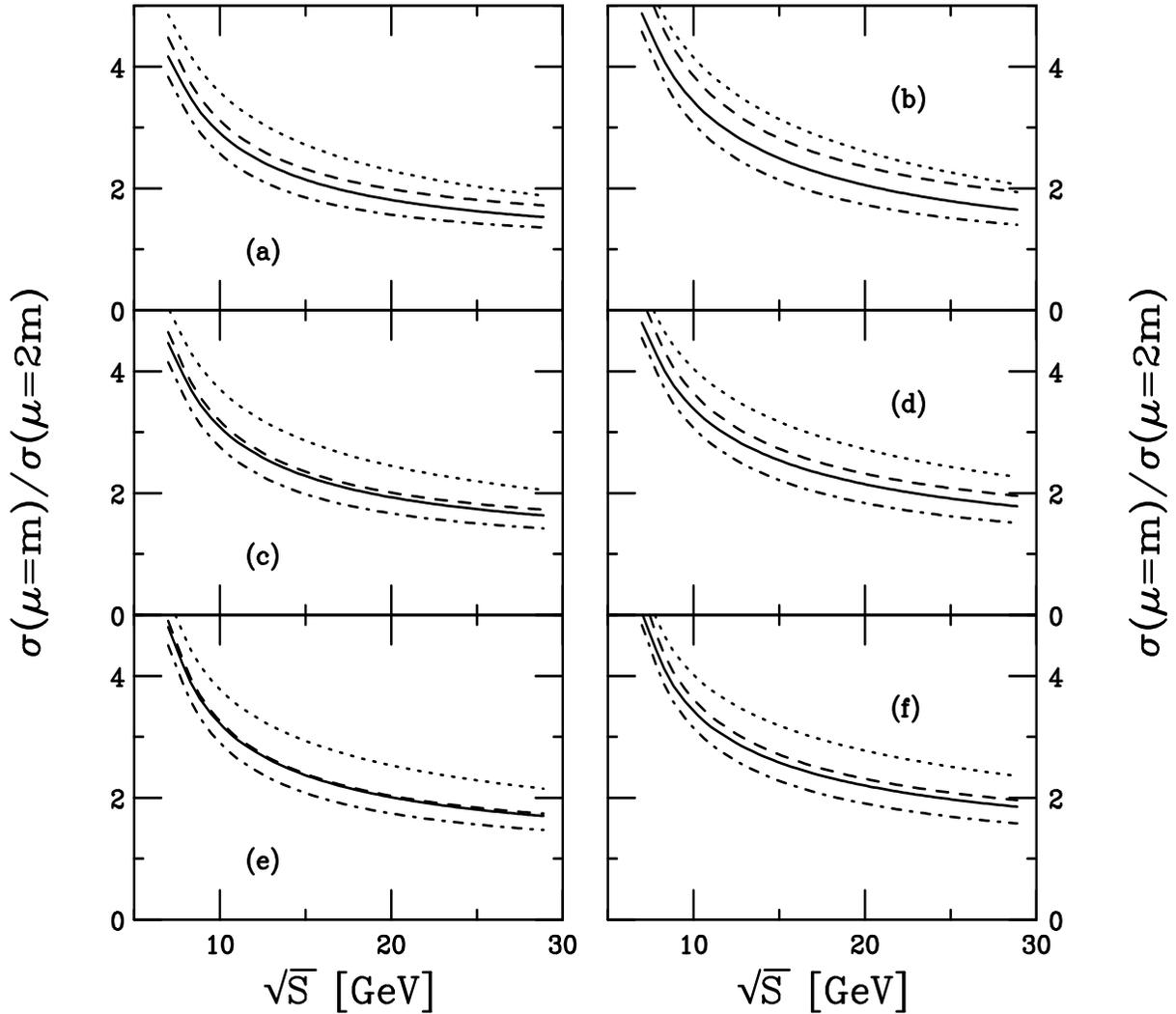}}
\caption[]{The scale dependence of $c{\bar c}$
production in $p p$ collisions with, (a) and (b), $m = 1.2$, (c) and (d), 
$m = 1.5,$ and, (e) and (f), $m = 1.8$ GeV.  We give the ratios
$\sigma(\mu = m)/\sigma(\mu = 2m)$ for the LO (dotted),
NLO (solid), 1PI NNLO-NNLL (dashed) and 1PI NNLO-NNNLL$+\zeta$ (dot-dashed) 
cross sections.  Results with the GRV98 HO densities are given on the 
left-hand side while the MRST2002 NNLO results are shown on the right-hand
side. 
}
\label{f15} 
\end{figure}

We compare the scale dependence of the cross sections in
Fig.~\ref{f15}.  The GRV98 HO cross section ratios on the left-hand side
are compared to the MRST2002 NNLO ratios on the right-hand side.  The scale
dependence is similar for the two sets of parton densities although the MRST
scale dependence is somewhat stronger at low $\sqrt{S}$.  The 1PI
NNLO-NNNLL$+\zeta$ scale dependence is reduced relative to the 1PI NNLO-NNLL
dependence which is stronger than the exact NLO dependence for most values of
mass.  Indeed the 1PI NNLO-NNLL ratios are only similar to the NLO ratios for
$m = 1.8$ GeV while the 1PI NNLO-NNNLL$+\zeta$ ratios are lower than the rest
of the calculated ratios for all masses.

\begin{figure}[htpb] 
\setlength{\epsfxsize=0.8\textwidth}
\setlength{\epsfysize=0.5\textheight}
\centerline{\epsffile{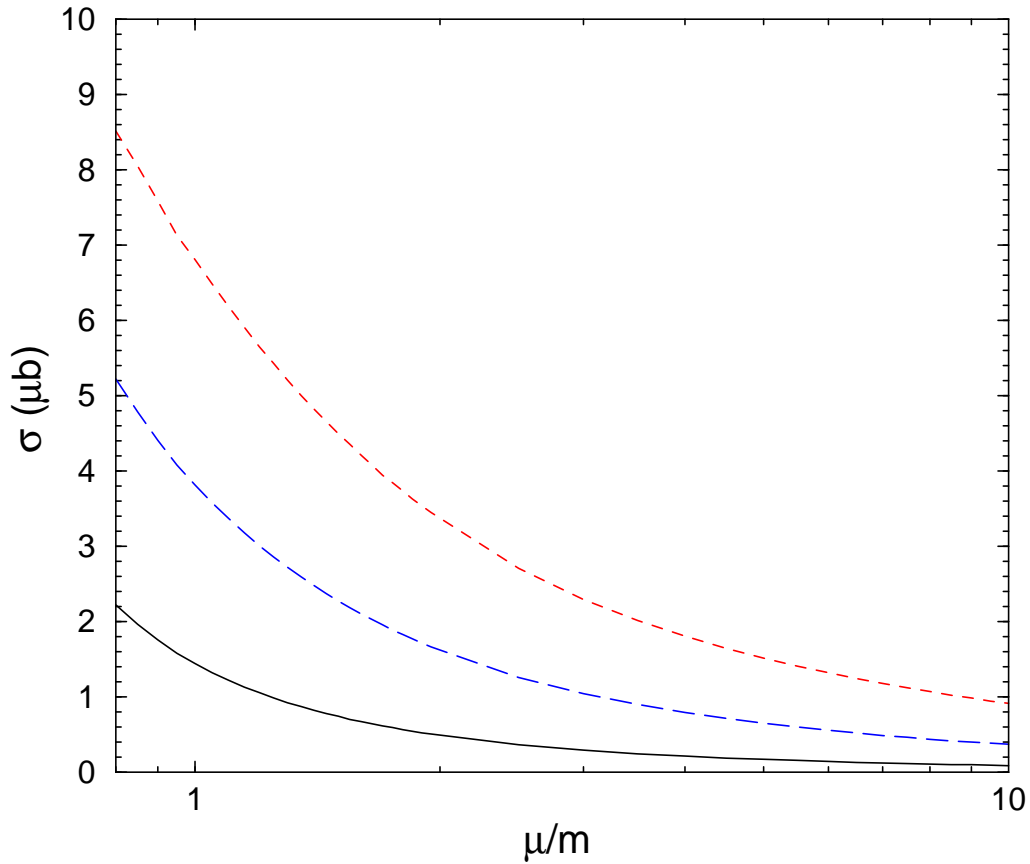}}
\caption[]{The charm production cross section as a function of $\mu/m$
at $\sqrt{S}=17.3$ GeV with $m = 1.5$ GeV and the MRST2002
NNLO parton densities. We show the 
Born (solid), NLO (long-dashed) and NNLO-NNNLL+$\zeta$ (dashed) results.
}
\label{charmmu} 
\end{figure}

In Fig.~\ref{charmmu} we plot the scale dependence for $0.8 < \mu/m < 10$ 
at $\sqrt{S}=17.3$ GeV and $m=1.5$ GeV. We show results for the
Born, NLO, and NNLO-NNNLL+$\zeta$ cross sections.  None of the cross sections
show a plateau at any value of $\mu/m$, as seen for the NLO and NNLO $b
\overline b$ cross sections in Fig.~\ref{botmu}.  However, the relative
dependence of the cross section on $\mu/m$ decreases as the accuracy of the
cross section increases.

\begin{figure}[htpb] 
\epsfxsize=0.5\textwidth\epsffile{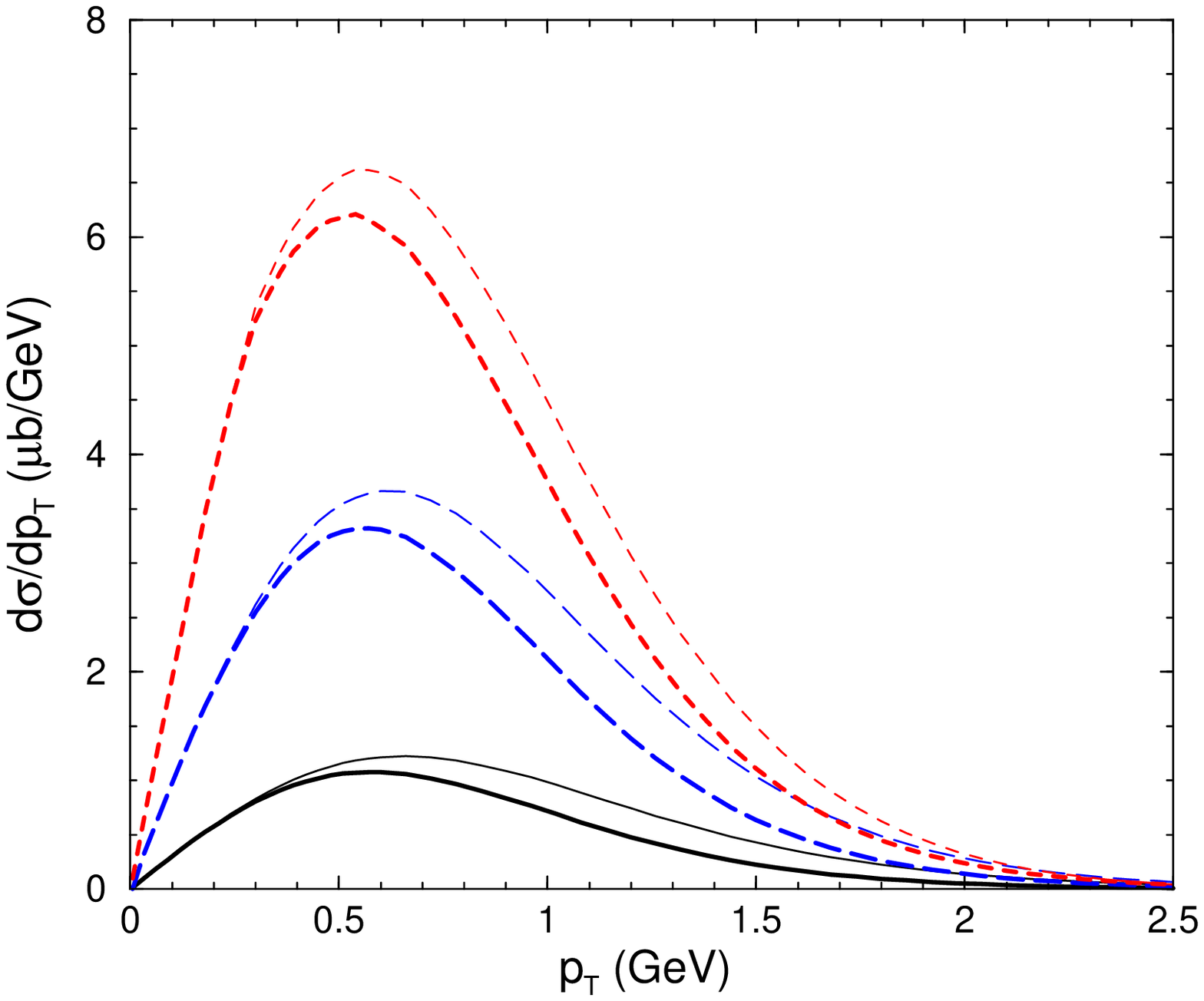}
\epsfxsize=0.5\textwidth\epsffile{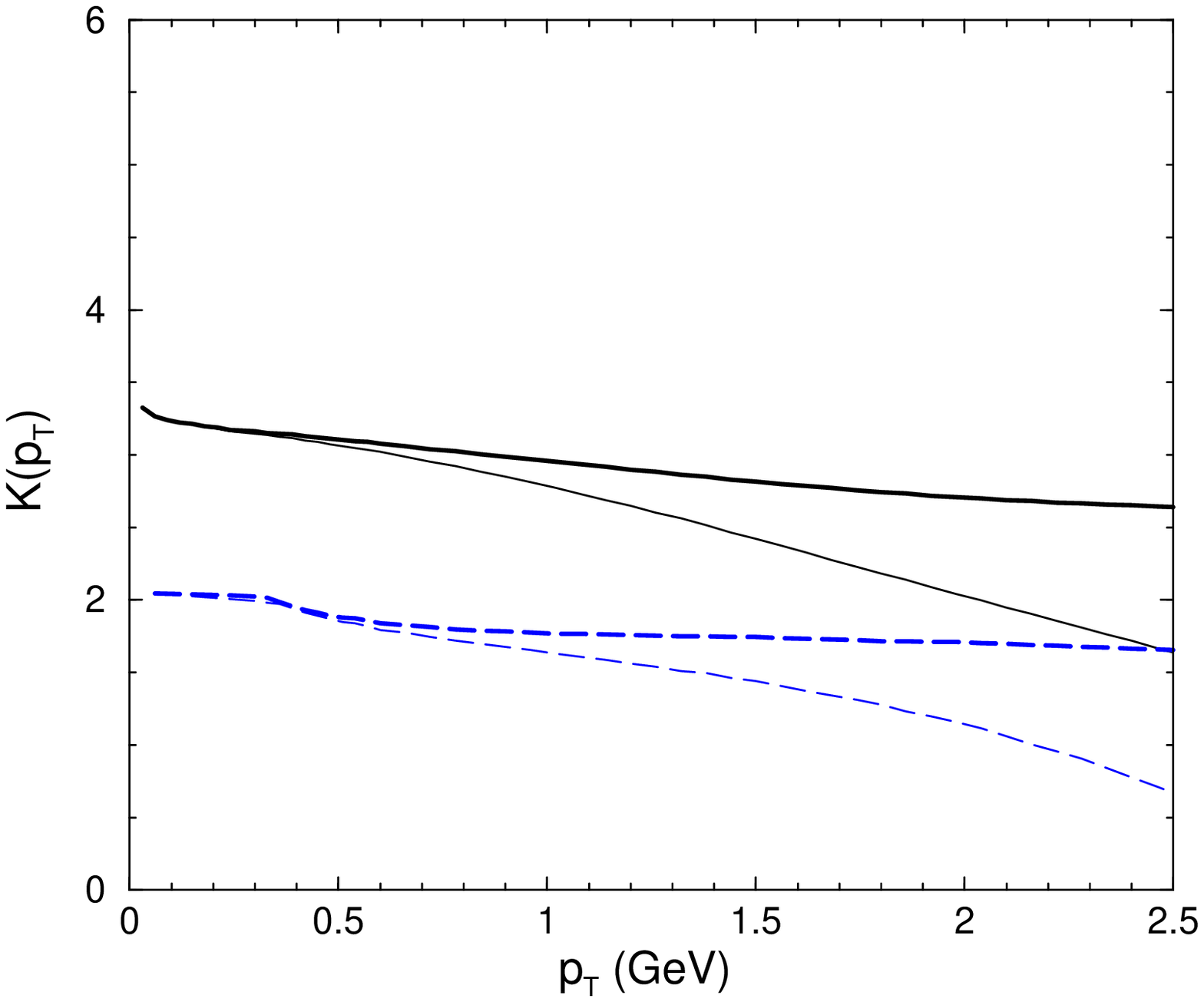}
\caption[]{The charm quark transverse momentum distributions at $\sqrt{S} =
17.3$ GeV and $m = 1.5$ GeV with the MRST2002 NNLO parton densities.  
On the left-hand side we show the Born (solid), NLO (long-dashed), 
and NNLO-NNNLL+$\zeta$ (dashed) results while on the right-hand side we
present $K_0^{(1)}$ (solid) and $K_{\rm sub}^{(2)}$ (dashed).  
The bold lines are calculated
with $\mu = m_T$ while the thin lines are with $\mu = m$.
}
\label{chmpt} 
\end{figure}

Figure \ref{chmpt} shows the charm quark transverse momentum distributions 
in $pp$ collisions at $\sqrt{S}=17.3$ GeV, $m = 1.5$ GeV, and two
scales, $\mu = m$ and $\mu=m_T=\sqrt{p_T^2+m^2}$. 
The Born, NLO, and NNLO-NNNLL+$\zeta$
results are shown on the left-hand side. 
On the right-hand side we plot $K_0^{(1)}$ and $K_{\rm sub}^{(2)}$.
Again the enhancement in the $p_T$ distribution is similar to that seen for
the total cross sections in Fig.~\ref{f14}.  The behavior of the $K$ factors as
a function of $p_T$ is similar to that in Fig.~\ref{botpt}.

\mysection{Conclusions}

In this paper we have calculated soft NNLO corrections
to the bottom and charm quark total cross sections and 
transverse momentum distributions
in hadron-hadron collisions. We have added new subleading soft NNNLL terms
and some virtual terms, including all soft-plus-virtual
factorization and renormalization scale-dependent terms. 
We have found that these new subleading corrections reduce the size of the NNLO
cross sections, and thus the $K$ factors, as well as
diminish the scale dependence of the cross section.

\mysection*{Acknowledgments}

The research of N.K. has been supported by a Marie Curie Fellowship of 
the European Community programme ``Improving Human Research Potential'' 
under contract number HPMF-CT-2001-01221.
The research of R.V. is supported in part by the 
Division of Nuclear Physics of the Office of High Energy and Nuclear Physics
of the U.S. Department of Energy under Contract No. DE-AC-03-76SF00098.

\end{document}